\documentclass [prb,twocolumn, amsmath,amssymb]{revtex4}
\setcitestyle{numbers,square}
\usepackage{appendix}
\usepackage{array}
\usepackage{amsmath,amssymb}
\usepackage{tabularx}
\usepackage{graphicx}
\usepackage{bmpsize}
\usepackage{bm}
\usepackage{color}
\usepackage{amssymb}
\usepackage [version=3]{mhchem}
\usepackage{newtxmath}
\usepackage{hyperref}
\usepackage{url}
\usepackage{physics}
\usepackage{here}
\usepackage{float}
\usepackage{comment}

\DeclareMathAlphabet{\mathpzc}{OT1}{pzc}{m}{it}

\def\para{\parallel}

\begin{document}

\title{
Perpendicular electronic transport and moir\'{e}-induced resonance in twisted interfaces of three-dimensional graphite
}
\author{Tenta Tani}
\affiliation{Department of Physics, Osaka University, Toyonaka, Osaka 560-0043, Japan}
\author{Takuto Kawakami}
\affiliation{Department of Physics, Osaka University, Toyonaka, Osaka 560-0043, Japan}
\author{Mikito Koshino}
\affiliation{Department of Physics, Osaka University, Toyonaka, Osaka 560-0043, Japan}
\date{\today}

\begin{abstract}
We calculate the perpendicular electrical conductivity in twisted three-dimensional graphite 
(rotationally stacked graphite pieces) by using the effective continuum model and the recursive Green's function method.
In the low twist angle regime $(\theta \lesssim 2^\circ)$, the conductivity shows a nonmonotonic dependence with a peak and dip structure as a function of the twist angle. By analyzing the momentum-resolved conductance and the local density of states, this behavior is attributed to the Fano resonance between continuum states of bulk graphite and interface-localized states, which is a remnant of the flat band in the magic-angle twisted bilayer graphene.
We also apply the formulation to the high-angle regime near the commensurate angle $\theta \approx 21.8^\circ$, and reproduce the conductance peak observed in the experiment.
%
\end{abstract}

\maketitle

\section{Introduction}
\label{sec:intro}
\begin{figure}
\begin{center}
   \includegraphics [width=0.6\linewidth]{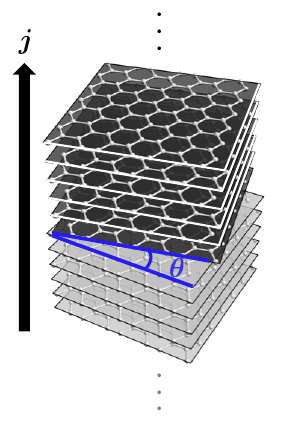}
   \caption{
   Structure of twisted graphite with a twist angle $\theta$.
   The black arrow indicates the perpendicular (out-of-plane) electronic transport.
            }\label{fig:twistedgr_schematic}
 \end{center}
 \end{figure}

In recent years, the field of twisted two-dimensional (2D) materials has attracted attention due to their unique and tunable physical properties. 
The concept of twisting 2D materials involves stacking two or more layers of the same or different materials with a specific twist angle between them. 
A representative system is twisted bilayer graphene (TBG), which consists of two graphene layers being rotated with respect to each other.
TBG hosts extremely flat bands at the Fermi energy at the so-called magic angle $\theta\approx1^\circ$~\cite{Bistritzer2011},
where various correlated phenomena have been experimentally observed~\cite{Cao2018SC,Cao2018Ins}.
Beyond TBG, the scope of research in the field has expanded to encompass twisted multilayers,
including twisted trilayer graphene~\cite{Eslam2019,Christophe2019,Carr2020,Hao2021,Xi2021,Park2021,Lei2021,Nakatsuji2023}, twisted double bilayer graphene (twist stack of two pieces of a Bernal-stacked bilayer) \cite{Koshino2019TDBG,Burg2019,Shen2020,Liu2020,Haddadi2020,Culchac2020,He2021,Szentpeteri2021,Petar2022} and twisted monolayer-bilayer graphene (monolayer and Bernal-stacked bilayer) \cite{Morell2013,Park2020,Louk2020,Chen2021,Li2022,Tong2022}.
In addition, research on twisted multilayer graphenes composed of a more general number of layers and configurations has also been conducted~\cite{Wu2014,Wu2015,Cea2019,Liu2019,Tritsaris2020,Nguyen2022,Park2022,Jie2022,Ledwith2022,Zhang2023,Waters2023}.
These systems often exhibit flat bands and associated peculiar physical properties.

This paper aims to extend the exploration to twisted three-dimensional (3D) systems where 3D layered materials are rotationally stacked
as shown in Fig.~\ref{fig:twistedgr_schematic}.
In particular, we focus on the twist-angle-dependent transport in the out-of-plane (perpendicular) direction, to explore measurable properties associated with the moir\'e pattern.
The conduction across twisted interfaces has been studied for various tunneling junctions with an insulating barrier in the middle, such as the graphene/hexagonal boron nitride/graphene structure \cite{Britnell2012nanolett,Britnell2012,Britnell2013,Kuzmina2021,Mishchenko2014,Ghazaryan2021,Seo2022}.
In these systems, the transport through the junction can be captured by a conventional perturbation approach including the tunneling process in the leading order \cite{Britnell2013,Kuzmina2021,Koprivica2022}.

When two materials are directly contacted, however, the multiple scattering at the twisted interface is generally relevant.
Here we consider a twisted 3D graphite (Fig.~\ref{fig:twistedgr_schematic}) as the simplest example of directly contacted twisted 3D systems.
The electronic structure of the twisted 3D graphite was previously studied, where a remnant of the flat band in TBG was found in the local density of states~\cite{Cea2019}.
The interlayer transport in twisted graphitic systems was investigated in various theoretical approaches~\cite{Bistritzer2010,Perebeinos2012,Ahsan2013,Fang2021,FANG2023}, and it was also experimentally probed in angle-variable devices \cite{Koren2016,Li2018,Inbar2023,Chari2016,Kim2013,Yu2020,Zhang2020-TBG}.
In large twist angles ($\theta\gtrsim 10^\circ$), it was predicted that the perpendicular conductance is enhanced near the commensurate angles where the atomic structure becomes exactly periodic \cite{Bistritzer2010}, and it was actually observed in conductance measurements as a sharp conductance peak against the twist angle \cite{Chari2016,Koren2016,Inbar2023}.
In this regime, the tunneling probability is small and the leading-order approximation in the transport is still valid.

In the present paper, we focus on the low twist angle regime where the multiple-scattering event is dominant.
We find a special resonant behavior in the  electronic transport due to the interface-localized states, which corresponds to the moir\'e flat band in TBG.
Specifically, we calculate the perpendicular electrical conductivity in twisted 3D graphite by combining the effective continuum model and the recursive Green's function method \cite{Soukoulis1982,MACKINNON1984,Ando1991},
to properly treat higher-order terms in the transmission.
In the low twist angle regime $(\theta \lesssim 2^\circ)$, in particular, we find that the perpendicular conductivity 
exhibits a peak-and-dip structure as a function of the twist angle. By analyzing the momentum-resolved conductance and the local density of states,
we attribute the sharp rise and drop of the conductivity to a Fano resonance between bulk states and the interface-localized states.
For the graphite band model,
we adopt a simplified circularly symmetric model as well as a more realistic version with the Slonczewski-Weiss-McClure (SWM) parameters fully included~\cite{McClure1957,Slonczewski1958}.
We confirm that the qualitative result does not depend on the choice of the models.
In the latter part of the paper, we apply the formulation to the high-angle regime near the commensurate angle $\theta \approx 21.8^\circ$, and simulate the conductance peak observed in the experiment \cite{Chari2016,Koren2016,Inbar2023}.
The formulation is applicable to general twisted 3D systems, such as 
a twisted interface of $\mathrm{NbSe_{2}}$ which exhibits the Josephson effect in the superconducting state~\cite{Yabuki2016}.

This paper is organized as follows.
In Sec.~\ref{sec:methods}, we introduce an effective continuum Hamiltonian of twisted graphite. 
In Sec.~\ref{sec:general}, we formulate the procedure to calculate the perpendicular electrical conductivity in twisted 3D systems using the recursive Green's function method and the effective continuum model.
In Sec.~\ref{sec:results}, we calculate the conductivity for 3D graphite with various twist angles, and find the non-monotonic behavior of the perpendicular conductivity.
In Sec.~\ref{sec:discussion}, we discuss the origin of the twist angle-dependence of the perpendicular conductivity.
We examine the momentum-resolved conductance and show that its sudden drop is due to the Fano resonance caused by the interface-localized state.
In Sec.~\ref{sec:commensurate}, we discuss the perpendicular conductivity near the second commensurate angle $\approx 21.8^\circ$.
Finally, the conclusion is given in Sec.~\ref{sec:conclusion}.

\section{Hamiltonian of twisted graphite}
\label{sec:methods}

\subsection{Band models for bulk graphite}
\label{subsec:bulk-graphite}

\begin{figure}
\begin{center}
   \includegraphics [width=0.7\linewidth]{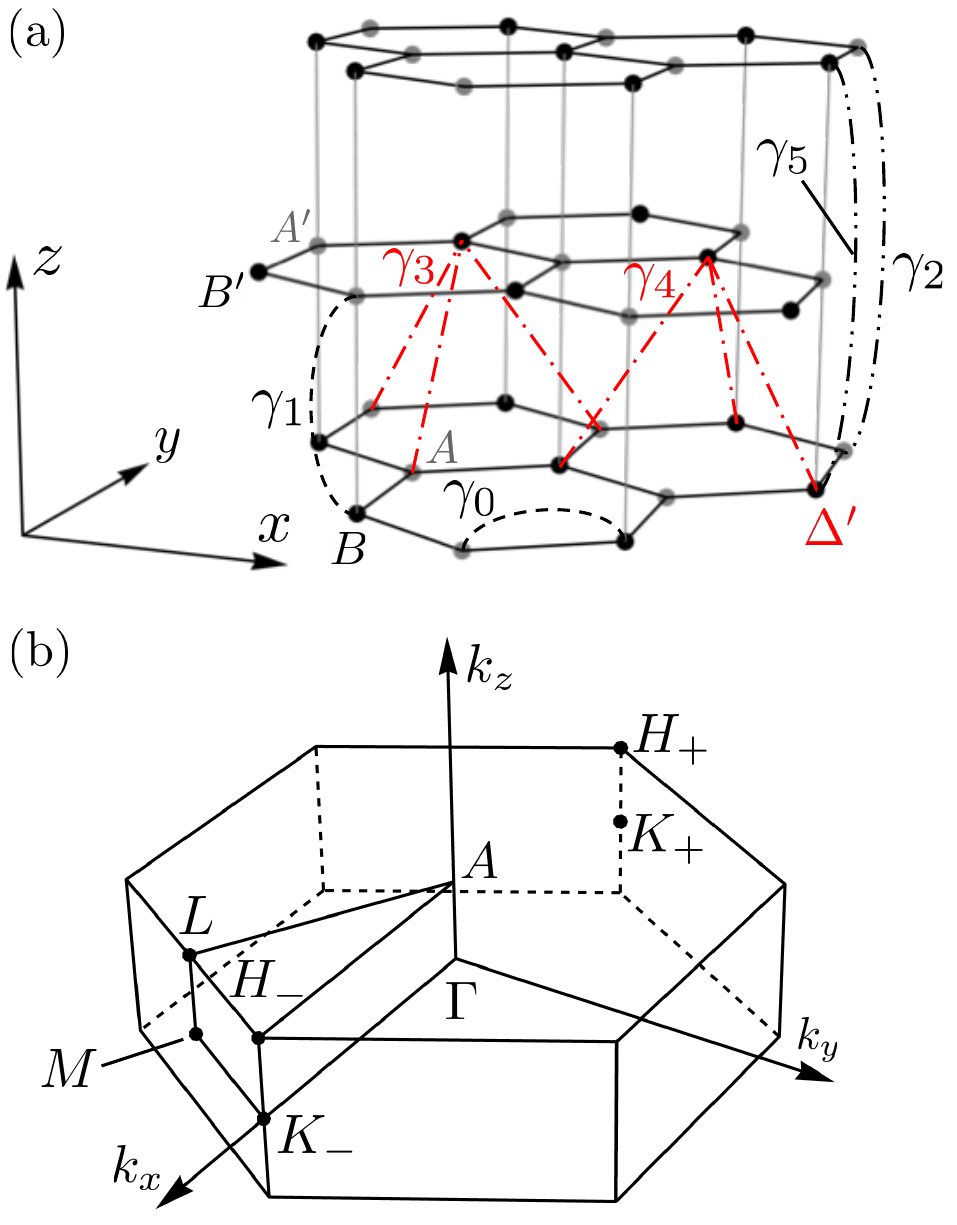}
   \caption{
   (a)~Crystal structure of graphite. 
  A unit cell spans over two layers, and contains four atomic sites $A,B,A',B'$.
   The hopping parameters are depicted by the dotted lines.
   The red lines ($\gamma_3, \gamma_4, \Delta'$) are present only in the full-parameter SWM model (see the text).
   (b)~Brillouin zone and high-symmetry points of graphite.          
}\label{fig:graphite_structure}
 \end{center}
 \end{figure}

{
The crystal structure of Bernal-stacking (AB-stacking) graphite is shown in Fig.~\ref{fig:graphite_structure}(a).
A unit cell contains four atomic sites 
$A,B,A',B'$,
where $B$ and $A'$ are arranged along vertical columns while $A$ and $B'$ are located above or below the center of hexagons in the neighboring layers.
The lattice constants are given by $a=0.246$~nm and $c=0.670$~nm for the in-plane and perpendicular direction, respectively.
We define primitive lattice vectors by $\vb*{a}_{1}=a(1,0,0)$, $\vb*{a}_{2}=a(1/2,\sqrt{3}/2,0)$ and $\vb*{a}_{3}=c(0,0,1)$.
The Brillouin zone is a hexagonal prism
spanned by the reciprocal lattice vectors
$\vb*{b}_{1}=(4\pi/\sqrt{3}a)(\sqrt{3}/2,-1/2,0)$, $\vb*{b}_{2}=(4\pi/\sqrt{3}a)(0,1,0)$, and $\vb*{b}_{3}=(2\pi/c)(0,0,1)$,
as shown in Fig.~\ref{fig:graphite_structure}(b).
The Fermi surfaces are located around $\vb*{K}_{\xi} = -(4\pi/3a)(\xi,0,0) \ (\xi=\pm1)$,
which are referred to as $K_{+}$ and $K_{-}$ points, respectively.
}


{We describe the electronic bands of graphite using the SWM model~\cite{McClure1957,Slonczewski1958,Koshino2009multilayer,mccann2013electronic}.
The model contains six hopping parameters $\gamma_{0},\gamma_{1},...,\gamma_{5}$ and an onsite energy $\Delta'$, which are visualized in Fig.~\ref{fig:graphite_structure}(a).
We use the values tabulated in Table \ref{tab:hoppings}.
\begin{table}
       \centering
        \caption{
        SWM hopping parameters used in the present paper (in units of eV).
        }
        \begin{tabularx}{85mm}{XXXXXXX}
    \hline \hline 
        $\gamma_{0}$ & $\gamma_{1}$ &  $\gamma_{2}$ & $\gamma_{3}$ & $\gamma_{4}$ & $\gamma_{5}$ & $\Delta'$ \\
    \hline
    $-2.47$ & $0.40$ & $-0.02$ & $0.30$ & $0.04$ & $0.04$ & $0.05$ \\
    \hline \hline
    \end{tabularx}
       \label{tab:hoppings}
   \end{table}
Taking the Bloch states 
$(\ket{A},\ket{B},\ket{A'},\ket{B'})$ as the basis,
the SWM Hamiltonian around the $K_{\xi}$ point is given by
\begin{equation}
    \begin{split}
         &H_{\mathrm{SWM}}(\vb*{k}_{\para},k_z) =
        H_\xi(\vb*{k}_{\para})
        +
        \left[T_\xi(\vb*{k}_{\para})e^{-ik_{z}c}
        +
        \text{h.c.}\right],
        \\
        &H_\xi(\vb*{k}_{\para}) =
        \mqty(0 & -\hbar v k_{-} & \hbar v_{4} k_{-} & \hbar v_{3} k_{+} \\
        -\hbar v k_{+} & \Delta' & \gamma_1 & \hbar v_{4} k_{-} \\
        \hbar v_{4} k_{+} & \gamma_1 & \Delta' & -\hbar v k_{-} \\
        \hbar v_{3} k_{-} & \hbar v_{4} k_{+} & -\hbar v k_{+} & 0
        ),
        \\
        &T_\xi(\vb*{k}_{\para}) = 
        \mqty(\gamma_{2}/2 & 0 & \hbar v_{4} k_{-} & \hbar v_{3} k_{+} \\
        0 & \gamma_{5}/2 & \gamma_1 & \hbar v_{4} k_{-} \\
        0 & 0 & \gamma_{5}/2 & 0 \\
        0 & 0 & 0 & \gamma_{2}/2
        ),
    \end{split}
    \label{eq:H_SWM}
\end{equation}
where $\vb*{k}_{\para}=(k_x,k_y)$ is the in-plane wavenumber measured from the $K_\xi$ point, $k_{\pm}=\xi k_x \pm ik_y$,
$v = \sqrt{3}|\gamma_0|a/2\hbar$, and $v_{i} = \sqrt{3}\gamma_{i}a/2\hbar \ (i=3,4)$.
In this paper, we consider a
simple model where we neglect $\gamma_3$, $\gamma_4$ and $\Delta'$, and the full-parameter model which contains all of the parameters.
}

\begin{figure*}
\begin{center}
   \includegraphics [width=0.7\linewidth]{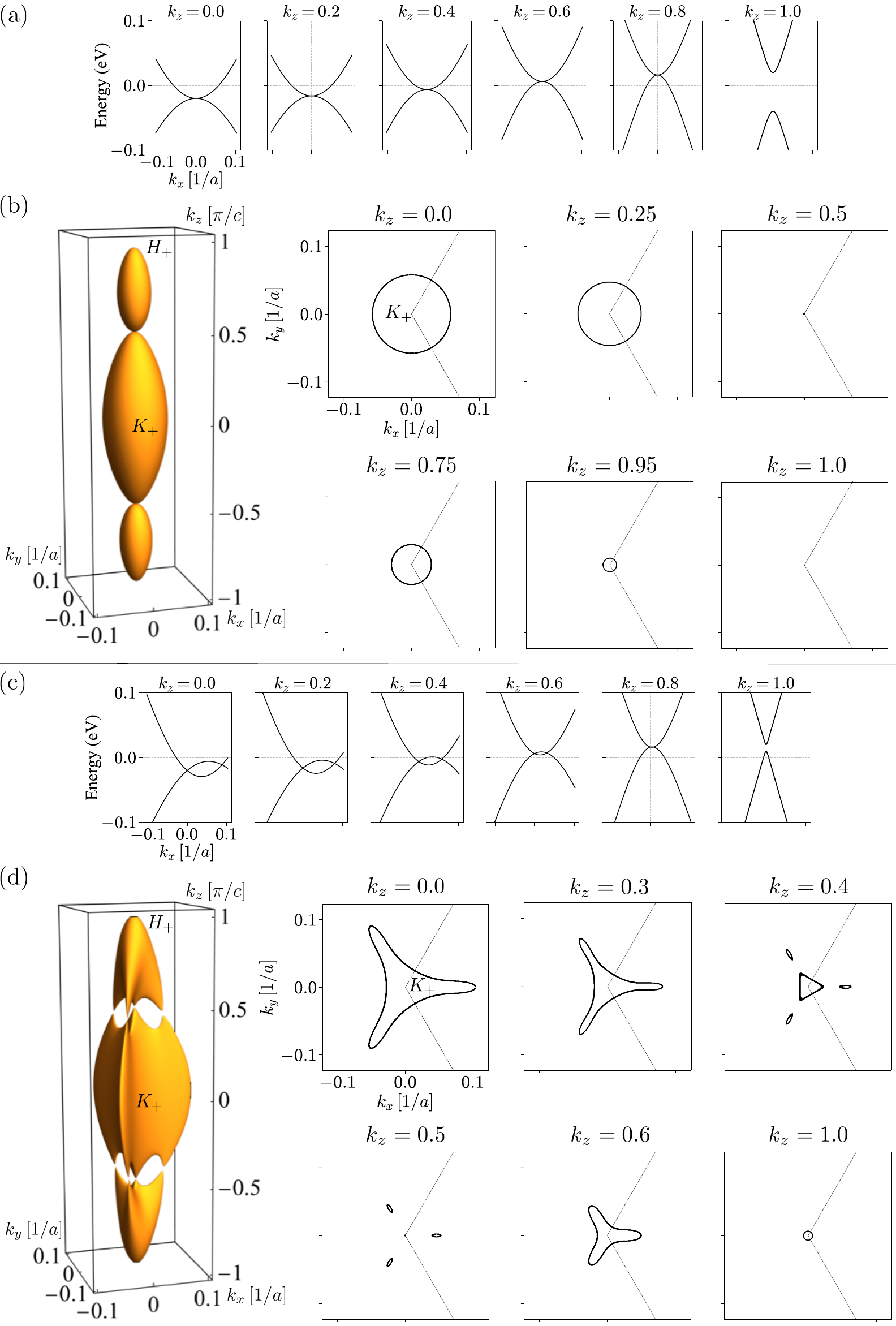}
   \caption{ 
   (a)~Band structure and (b)~the Fermi surface of the simple SWM model of graphite.
   Panel (a) shows the in-plane band dispersion with fixed $k_z$'s, which are indicated in units of $\pi/c$.
   In (b), the right panels indicate the cross sections at fixed $k_z$.
   (c) and (d) are the corresponding figures for the full-parameter SWM model.
   The origin of in-plane momentum $(k_x,k_y)$ is set to the $K_+$ point.
            }\label{fig:Fermisurface}
 \end{center}
 \end{figure*}

The band structures and Fermi surfaces for these two models are shown in Fig.~\ref{fig:Fermisurface}.
Due to the $\gamma_2$ parameter, the band structure obtains the dispersion along $k_z$, which gives rise to the formation of the electron and hole pocket.
Notably, the isotropic Fermi surface in the simple model is warped 
in a 120$^\circ$-symmetric manner
in the full-parameter model. 
This trigonal warping effect is due to the entry of the $\gamma_3$  parameter.

\subsection{Twisted graphite}
\label{subsec:conductivity_twistedgr}

\begin{figure}
\begin{center}
   \includegraphics [width=0.9\linewidth]{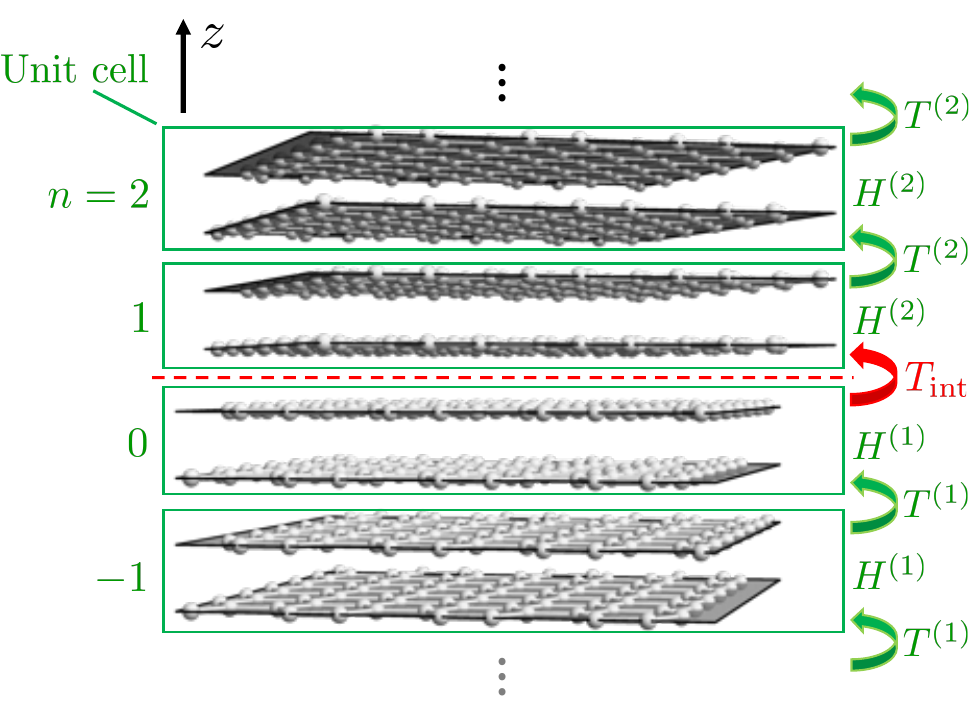}
   \caption{
   Side view of twisted graphite.
   The unit cells of bulk graphite are indicated by the green rectangles, which are labeled by integers $n$.
   The red dashed line laid between the $n=0$ and $n=1$ cell is the twisted interface.
   The lower (upper) bulk graphite is twisted by the angle $-\theta/2$ ($+\theta/2$) from the aligned position.
   The green and red arrows are the hopping matrices between nearest-neighbor unit cells.
            }\label{fig:twistedgr_structure}
 \end{center}
 \end{figure}

We define a twisted graphite as a pair of half-infinite pieces of Bernal-stacking graphite contacted with a certain twist angle $\theta$.
A schematic illustration is given in Fig.~\ref{fig:twistedgr_structure},
where we label unit cells of graphite (extending over two graphene layers) by $n = 0, \pm1, \pm2, \cdots$. Here $n>0$ and $n\leq 0$ correspond to upper and lower graphite pieces, respectively,
and we label sublattices in the $n$th cell as $A_n, B_n, A'_n, B'_n$.

\begin{figure}
\begin{center}
   \includegraphics [width=\linewidth]{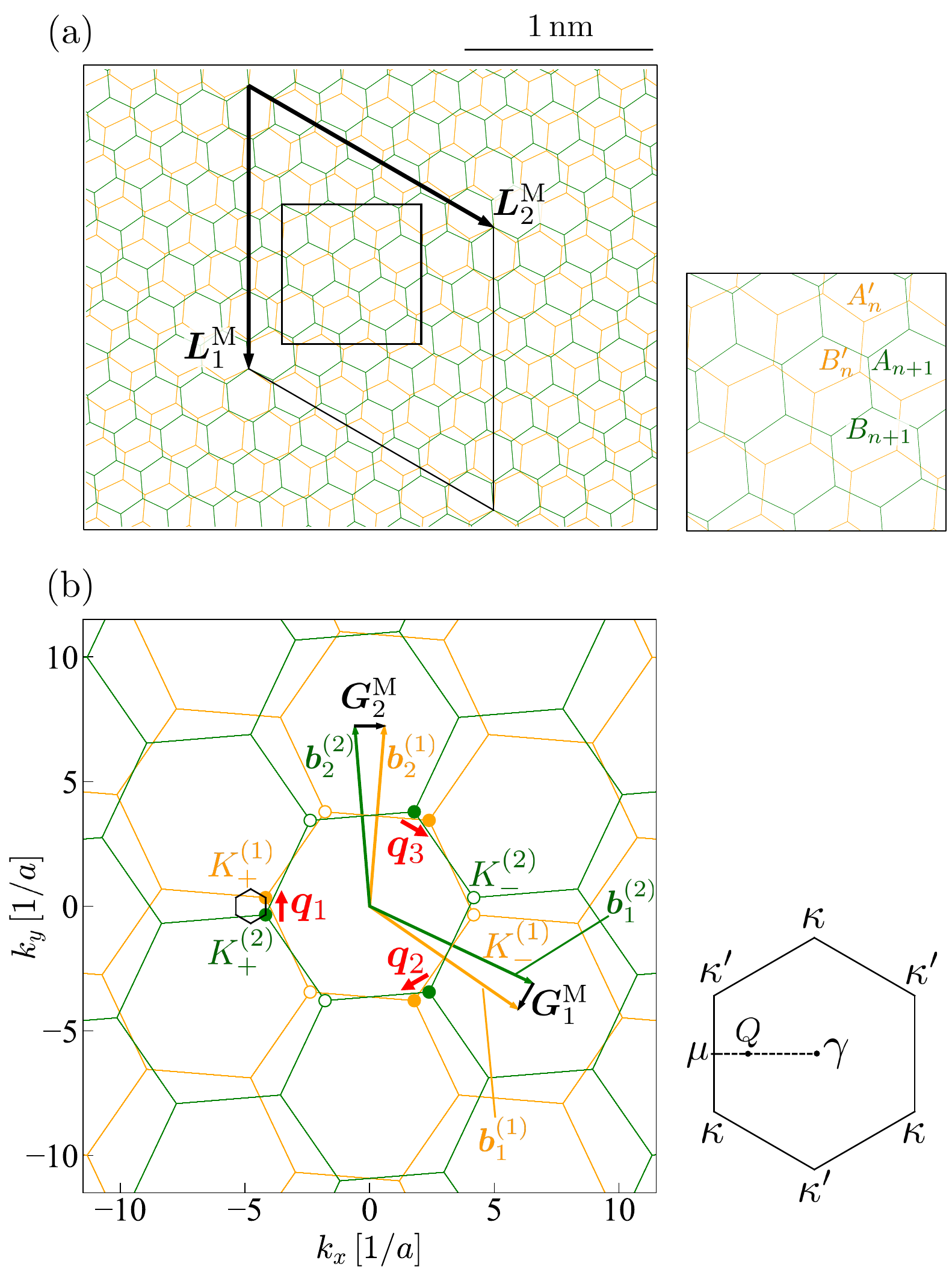}
   \caption{
   (a) Atomic structure of twisted graphite interface with
   $\theta=9.43^\circ$, consisting of the lower (yellow) and upper (green) honeycomb lattice.
   The black rhombus is a moir\'{e} unit cell.
 The right panel is a magnified plot with sublattice labels.
   (b) Brillouin zone of the lower (yellow) and upper (green) honeycomb lattice in the extended-zone scheme.
   The coupling wavenumbers $\vb*{q}_1,\vb*{q}_2,\text{and} \, \vb*{q}_3$ are shown by the red arrows (see the text).
  A moir\'{e} Brillouin zone is defined by a small black hexagon near $\vb*{q}_1$, which is magnified in
    the right figure with labels for high-symmetry points.
            }\label{fig:MBZ}
 \end{center}
 \end{figure}

At the interface, a long-scale moir\'{e} pattern is formed. 
Figure \ref{fig:MBZ}(a) illustrates the atomic structure of the twisted interface
between the $n$th and $(n+1)$th cells.
We define the primitive lattice vectors of the lower ($l=1$) and upper ($l=2$) graphite as 
$\bm{a}^{(l)}_i = R^{(l)}\bm{a}_{i}$,
and also the primitive reciprocal vectors as
$\bm{b}^{(l)}_{i} = R^{(l)}\bm{b}_{i}$,
where
\begin{equation}
    R^{(1)} = R(-\theta/2),\quad
    R^{(2)} = R(+\theta/2),
\end{equation}
and 
$R(\alpha)$ is the rotation matrix by an angle $\alpha$.
Accordingly, the corner points of the Brillouin zones 
are given by 
$\bm{K}_{\xi}^{(l)} = R^{(l)} \bm{K}_{\xi}$.

The moir\'{e} Brillouin zone (MBZ) is defined by the reciprocal vectors $\vb*{G}^{\mathrm{M}}_{i} = \vb*{b}^{(1)}_{i} - \vb*{b}^{(2)}_{i}$, as depicted in Fig.~\ref{fig:MBZ}(b).
We also introduce the displacement of the $K_+$ point as
$\bm{q}_1 = \bm{K}_{+}^{(1)} - \bm{K}_{+}^{(2)}$,
and also $\bm{q}_2 = R(120^\circ)\bm{q}_1,  \, \bm{q}_3= R(-120^\circ)\bm{q}_1$ for the other two equivalent corners. We have relationships
$\bm{G}^{\mathrm{M}}_1 = \bm{q}_2 - \bm{q}_1$ and 
$\bm{G}^{\mathrm{M}}_2 = \bm{q}_3 - \bm{q}_2$.
The primitive moir\'{e} lattice vector in the real space is determined by $\vb*{L}^{\mathrm{M}}_{i} \cdot \vb*{G}^{\mathrm{M}}_{j} = 2\pi\delta_{ij}$~[see Fig.~\ref{fig:MBZ}(a)], giving $\vb*{L}^{\mathrm{M}}_{i} = -\vb*{e}_{z}\cross\vb*{a}_{i}/2\sin{(\theta/2)}$ for $i=1,2$ \cite{Moon2013}.


We describe the electronic structure of twisted graphite by an effective continuum model similar to twisted bilayer graphene \cite{Bistritzer2011,Koshino2018Wannier}.
In a basis of 
$(\cdots;\ket{A_n},\ket{B_n},\ket{A'_n},\ket{B'_n};\ket{A_{n+1}},\ket{B_{n+1}},$ 
$\ket{A'_{n+1}},\ket{B'_{n+1}};\cdots)$,
the Hamiltonian of twisted graphite can be written as
\begin{equation}
\begin{split}
{\cal H} =
    \left(
    \begin{array}{cccc|cccc}
    \ddots & \ddots & \ddots & & & & & 
    \\
    & T^{(1)} & H^{(1)} & T^{(1)\dagger} & & & & 
    \\
    & & T^{(1)} & H^{(1)} & T_{\mathrm{int}}^{\dagger} & & &
    \\[2pt]
    \hline
    & & & T_{\mathrm{int}} & H^{(2)} & T^{(2)\dagger} & &
    \\
    & & & & T^{(2)} & H^{(2)} & T^{(2)\dagger} & 
    \\
    & & & & & \ddots & \ddots & \ddots 
    \\
    \end{array}\right),
\end{split}
\label{eq:overall-Hamil3}
\end{equation}
where 
$H^{(l)}$ and $T^{(l)}$ are $4\times 4$ blocks
with $l=1$ and 2 indicating the lower and upper graphite sectors, respectively, which are given by
\begin{align}
H^{(l)} = H_\xi[(R^{(l)})^{-1}\bm{k}_{\para}],
\quad
 T^{(l)} = T_\xi[(R^{(l)})^{-1}\bm{k}_{\para}].
\label{eq:twistedgr-Tmatrix}
\end{align}
Here $H$ and $T$ are defined in Eq.~\eqref{eq:H_SWM}.

$T_{\mathrm{int}}$ is the interlayer Hamiltonian matrix for the twisted interface, which is given by \cite{Bistritzer2011,Koshino2018Wannier}
\begin{equation}
    \begin{split}
        T_{\mathrm{int}} &= \mqty(0 & T_{\mathrm{int}}^{2\times 2} \\ 0 & 0),
        \\
        T_{\mathrm{int}}^{2\times 2} (\vb*{r}) &=
        \mqty(u&u'\\u'&u)  e^{i\xi\bm{q}_{1}\cdot\vb*{r}}
        + \mqty(u&u'\omega^{-\xi}\\u'\omega^{\xi}&u)
        e^{i\xi\bm{q}_{2}\cdot\vb*{r}}
        \\
        &\hspace{1.5cm} + \mqty(u&u'\omega^{\xi}\\u'\omega^{-\xi}&u)
        e^{i\xi\bm{q}_{3}\cdot\vb*{r}},
    \end{split}
\label{eq:interlayer}
\end{equation}
where $\omega=e^{2\pi i/3}$, $u=0.0797~\mathrm{eV}$, and $u'=0.0975~\mathrm{eV}$~\cite{Koshino2018Wannier}.
By starting from a lower-graphite Bloch state of the wavenumber $\vb*{k}_{\para}$, the interlayer Hamiltonian $T_{\mathrm{int}}$ hybridizes a 
set of wavenumbers of the same valley,
\begin{align}
\label{eq:q}
  &  \bm{k}^{(1)}_{\para}(m_1,m_2) = \bm{k}_{\para} + m_{1}\bm{G}^{\mathrm{M}}_{1} + m_{2}\vb*{G}^{\mathrm{M}}_{2},
    \nonumber\\
  &   \bm{k}^{(2)}_{\para}(m_1,m_2) = \xi\bm{q}_1 + \bm{k}_{\para} + m_{1}\bm{G}^{\mathrm{M}}_{1} + m_{2}\vb*{G}^{\mathrm{M}}_{2}
\end{align} 
of the upper and lower parts, respectively ($m_1, m_2$: integers).
To write down the Hamiltonian as a finite-sized matrix, we consider a finite set of wavenumbers
inside a certain cutoff circle $|\bm{k}^{(l)}_{\para}| \leq k_{\mathrm{c}}$.
Note that $\vb*{k}_{\para}$ is a parameter which moves inside a moir\'e Brillouin zone spanned by $\vb*{G}^{\mathrm{M}}_{1}$ and $\vb*{G}^{\mathrm{M}}_{2}$ [Fig.~\ref{fig:MBZ}(b)].
In this representation, $H^{(l)}$, $T^{(l)}$ and $T_{\mathrm{int}}$ in 
Eq.~\eqref{eq:overall-Hamil3}
are $4N^{(l)}_q\times 4N^{(l)}_q$ matrices, where
$N^{(l)}_q$ is the number of different wavenumbers in the set of $\{\bm{k}^{(l)}_{\para} \}$, and 
the factor 4 is for the sublattices ($A,B,A',B'$).
The matrices $H^{(l)}$ and $T^{(l)}$ are diagonal in the label $(m_1,m_2)$, while only the matrix $T_{\mathrm{int}}$ hybridizes different $(m_1,m_2)$’s.

\section{Electrical conductivity in general twisted interface}
\label{sec:general}

{
In this section, we formulate a method to calculate the perpendicular electrical conductivity and the local density of states (LDOS) in general twisted 3D systems.
We consider a layered system as shown in Fig.~\ref{fig:3d_system_structure},
which consists of slices labeled by indices $n$.
A slice can be a single atomic layer or a cluster of layers.
The whole system is composed of the
lower ($n\leq0$), middle ($1\leq n\leq N$), and upper ($n\geq N+1$) parts. We assume that the lower and upper parts are periodic in the $z$ direction (perpendicular to the layer) with a single period corresponding to a single $n$.
The middle part can be periodic or nonperiodic in the $z$ direction,
and they can be arranged in a general orientation.
We require that all the layers in the upper, middle and lower parts share a common super-periodicity in in-plane directions (e.g., the moir\'e period for the twisted graphite),
so that the Hamiltonian becomes a finite matrix in a momentum representation, under a certain $k$-space cutoff.
The twisted graphite corresponds to 
a system without a middle part $(N=0)$, while we can formally assign the middle part within the same system by taking an arbitrary number of upper and lower layers including the twisted interface.
We apply the formulation with $N=0$ to the calculation of the conductivity, while a formalism with a finite middle part can be used to calculate the LDOS near the interface, where the middle part is set to contain the desired region. 
%
}



{
In a similar manner to the twisted graphite in the previous section, the Hamiltonian of the system is written as
\begin{widetext}
\begin{equation}
    {\cal H} = 
    \\
    \left(
    \begin{array}{cccc|cccc|cccc}
    \ddots & \ddots & \ddots & & & & & & & & &
    \\
    & T^{(1)} & H^{(1)} & T^{(1)\dagger} & & & & & & & &
    \\
    & & T^{(1)} & H^{(1)} & T_{10}^{\dagger} & & & & & & &
    \\
    [2pt]
    \hline
    & & & T_{10} & h_1 & T_{21}^{\dagger} & & & & & &
    \\
    & & & & T_{21} & h_2 & \ddots & & & & &
    \\
    & & & & & \ddots & \ddots & T_{N,N-1}^{\dagger} & & & & 
    \\
    & & & & & & T_{N,N-1} & h_N & T_{N+1,N}^{\dagger} & & &
    \\
    [2pt]
    \hline
    & & & & & & & T_{N+1,N} & H^{(2)} & T^{(2)\dagger} & 
    \\
    & & & & & & & & T^{(2)} & H^{(2)} & T^{(2)\dagger}
    \\
    & & & & & & & & & \ddots & \ddots & \ddots 
    \\
    \end{array}\right),
\label{eq:overall-Hamil1}
\end{equation}
\end{widetext}
where $H^{(l)}$ represents the Hamiltonian of a single slice in the lower ($l=1$) and upper ($l=2$) regions,
and $T^{(l)}$ is the hopping matrix between neighboring slices.
$H^{(l)}$ and $T^{(l)}$ are $M_l\times M_l$ matrices. 
Here, we ignore hopping terms across more than two slices. This is justified by taking sufficiently large slices.
$h_n$ and $T_{n,n+1}$ are intra- and interslice matrices, respectively, 
in the middle part.
The dimension of $h_n$ is arbitrary.
The total Hamiltonian is labeled by
$\vb*{k}_{\para}$ in the Brillouin zone
corresponding to the in-plane supercell of the system.
}


{
The electrical conductance in the perpendicular direction can be calculated by applying the recursive Green's function method~\cite{Ando1991,Lewenkopf2013} to the Hamiltonian Eq.~\eqref{eq:reduced_hamil}.
Specifically, we calculate eigenchannels of the upper and lower periodic parts, and express transmission coefficients between these channels using the Green's function, as follows.
}

{
To obtain the eigenchannels, we solve
the Schr\"odinger equation for the upper and lower periodic parts
\begin{equation}
    E \vb*{C}_{n} = H^{(l)} \vb*{C}_{n}
    + T^{(l)\dagger} \vb*{C}_{n+1}
    + T^{(l)} \vb*{C}_{n-1},
\end{equation}
where $E$ is the eigenenergy
and $\vb*{C}_{n}$ is $M_l$-component vector.
We first assume a solution of Bloch's form
$ \vb*{C}_{n} =  \lambda^n \vb*{C}_{0}$.
By using $ \vb*{C}_{n+1} =  \lambda \vb*{C}_{n}$,
we obtain
\begin{equation}
    \mqty([T^{(l)\dagger}]^{-1}(E-H^{(l)}) & -[T^{(l)\dagger}]^{-1}T^{(l)} \\ 1 & 0)
    \mqty(\vb*{C}_{n}\\ \vb*{C}_{n-1})
    =
    \lambda 
    \mqty(\vb*{C}_{n}\\ \vb*{C}_{n-1}).
\label{eq:non-hermitian}
\end{equation}
Equation~(\ref{eq:non-hermitian}) can be viewed as a $2M_l \times 2M_l$ eigenvalue problem with an eigenvalue $\lambda$.
For a given energy $E$, 
we obtain $M_l$ upward-going solutions
$\vb*{C}_0 = \vb*{u}^{(l)}_{1,+},\cdots,\vb*{u}^{(l)}_{M_l,+}$
with eigenvalues
$\lambda^{(l)}_{1,+},\cdots,\lambda^{(l)}_{M_l,+}$,
and 
$M_l$ downward-going solutions
$\vb*{u}^{(l)}_{1,-},\cdots,\vb*{u}^{(l)}_{M_l,-}$
with 
$\lambda^{(l)}_{1,-},\cdots,\lambda^{(l)}_{M_l,-}$.
Here upward- (downward-) going solutions include propagating modes in the positive (negative) $z$ direction and evanescent modes decaying in the positive (negative) $z$ direction.
%

\begin{figure}[b]
\begin{center}
   \includegraphics [width=0.8\columnwidth]{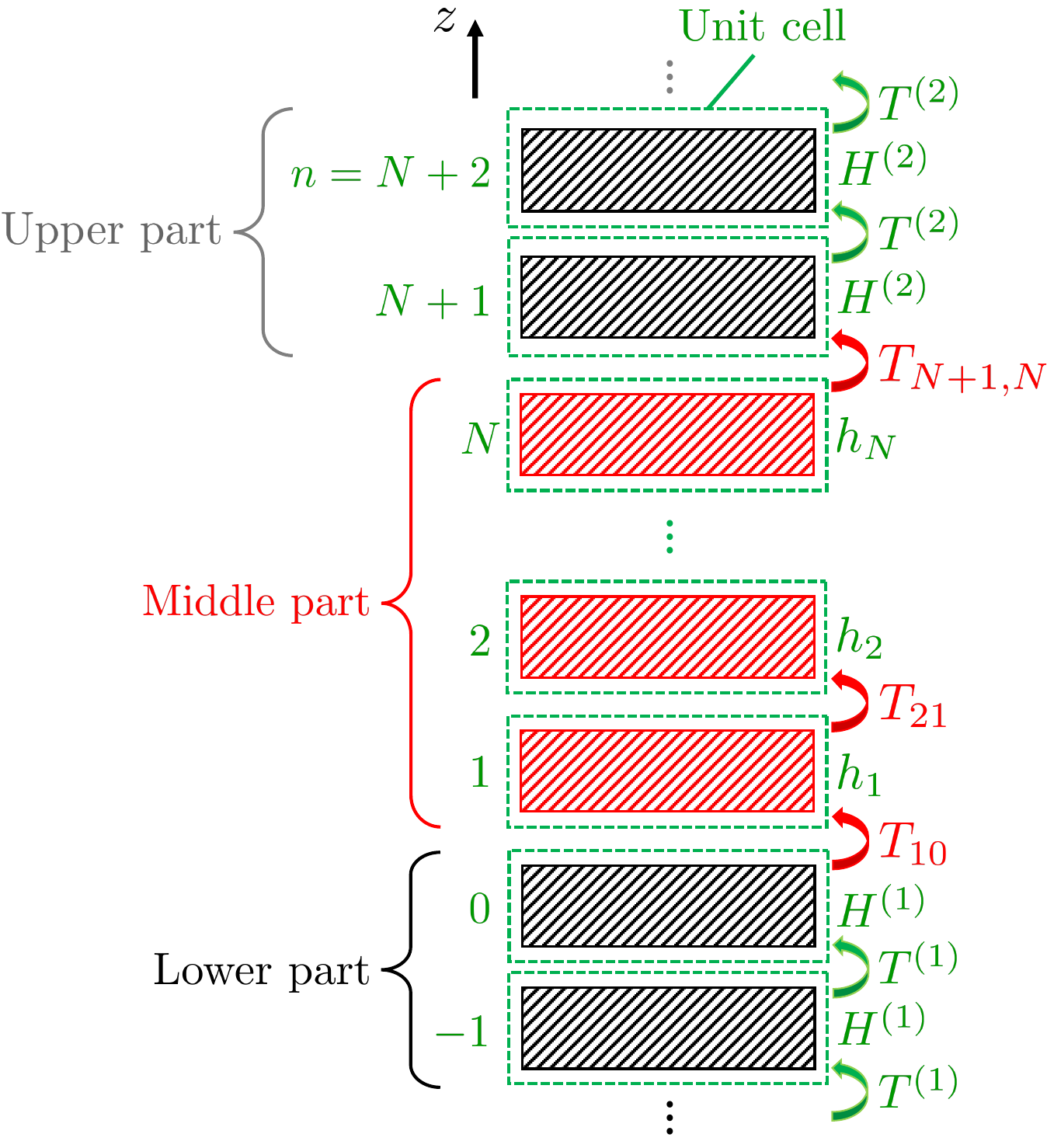}
   \caption{
   Schematic picture of a general twisted 3D system.
The structure consists of slices labeled by $n$, where
each slice can be a single atomic layer or a cluster of layers.
The lower part ($n\leq0$) and upper part ($n\geq N+1$) are periodic in $z$ direction, while
the middle part ($1\leq n\leq N$) can be arranged in a general orientation.
}\label{fig:3d_system_structure}
 \end{center}
 \end{figure}

A general solution at $n=0$ can be written in a linear combination of these eigenfunctions as
\begin{align}
& \vb*{C}_0 = \vb*{C}^{(+)}_0  + \vb*{C}^{(-)}_0,
\nonumber\\
& \vb*{C}^{(\pm)}_0  =  
{c}^{(\pm)}_1 \vb*{u}^{(l)}_{1,\pm}
+
{c}^{(\pm)}_2 \vb*{u}^{(l)}_{2,\pm}
+ \cdots
+ {c}^{(\pm)}_{M_l} \vb*{u}^{(l)}_{M_l,\pm}.
\end{align}
The wavefunction at general positions can be found by  
\begin{align}
& \vb*{C}^{(\pm)}_{n+1} = F^{(l)}_{\pm} \vb*{C}^{(\pm)}_{n},
\end{align}
where 
\begin{equation}
\begin{split}
 F^{(l)}_{\pm} &=
    U^{(l)}_{\pm}
    \Lambda^{(l)}_{\pm}
    \left(U^{(l)}_{\pm}\right)^{-1},
\\
    U^{(l)}_{\pm} &= 
        \left(
        \vb*{u}^{(l)}_{1,\pm},
        \vb*{u}^{(l)}_{2,\pm},\cdots,
        \vb*{u}^{(l)}_{M_l,\pm}
        \right),
    \\
    \Lambda^{(l)}_{\pm} &=
    \mathrm{diag}
    \left(
    \lambda^{(l)}_{1,\pm},
    \lambda^{(l)}_{2,\pm},\cdots,
    \lambda^{(l)}_{M_l,\pm}
    \right)
\end{split}
\end{equation}
are $M_{l}\times M_{l}$ matrices.
}

{
An eigenvalue equation for the middle part is then written as
\cite{Ando1991} 
\begin{widetext}
\begin{align}
&   (E - \mathcal{H}_{\mathrm{reduced}})
    \mqty(\vb*{C}_{0} \\ \vb*{C}_{1} \\ \vdots \\ \vb*{C}_{N+1})
    =
    \mqty(
    T^{(1)}
    \left( [F_{+}^{(1)}]^{-1} - [F_{-}^{(1)}]^{-1} \right)
    \vb*{C}_{0}^{(+)}
        \\ 0 \\ \vdots \\ 0
        ),
        \label{eq:source_equation}
\\
&    \mathcal{H}_{\mathrm{reduced}}=   
    \left(
    \begin{array}{c|cccc|c}
    H^{(1)}+\Sigma^{(1)} & T_{10}^{\dagger} & & & &
    \\
    [2pt]
    \hline
    T_{10} & h_1 & T_{21}^{\dagger} & & & 
    \\
    & T_{21} & h_2 & \ddots & & 
    \\
    & & \ddots & \ddots & T_{N,N-1}^{\dagger} &
    \\
    & & & T_{N,N-1} & h_N & T_{N+1,N}^{\dagger} 
    \\
    [2pt]
    \hline
    & & & & T_{N+1,N} & H^{(2)}+\Sigma^{(2)}
    \end{array}\right),
\label{eq:reduced_hamil}
\end{align}
\end{widetext}
where $\Sigma^{(1)}$ ($\Sigma^{(2)}$) represents the self energy matrix for the open leads in the lower (upper) part,
which are defined by
\begin{equation}
    \begin{split}
        \Sigma^{(1)} &= T^{(1)}[F^{(1)}_{-}]^{-1},
        \\
        \Sigma^{(2)} &= T^{(2)\dagger}F^{(2)}_{+}.
    \end{split}
\end{equation}
{The term with $\vb*{C}_{0}^{(+)}$ on the right-hand side of Eq.~\eqref{eq:source_equation} represents a source term associated with the incident wave from the lower channels.}
The Green's function of the system is defined by $\mathscr{G} = (E-\mathcal{H}_{\mathrm{reduced}})^{-1}$.
To obtain this, the recursive method can be utilized.
The details of the method are explained in Appendix~\ref{app:recursive}.
}

{
Once we have the Green's function $\mathscr{G}$, 
the transmission coefficients $t_{\mu\nu}$ are obtained by
\begin{equation}
    \begin{split}
        & t_{\mu\nu} = \sqrt{\frac{v_{\mu,+}^{(2)}}{v_{\nu,+}^{(1)}}} \ \times
        \\
        &
        \left[
        [U_{+}^{(2)}]^{-1}
        \mathscr{G}_{N+1,0}
        T^{(1)}
        \left([F_{+}^{(1)}]^{-1} - [F_{-}^{(1)}]^{-1}\right)
        U_{+}^{(1)}
        \right]_{\mu\nu},
    \end{split}
\label{eq:transmission_coeff}
\end{equation}
where $v^{(1)}_{\mu,+} (v^{(2)}_{\mu,+})$ is the group velocity of the $\mu$-th upward eigenmodes in the lower (upper) bulk.
The $M_2\times M_1$ matrix $\mathscr{G}_{N+1,0}$ is the partial block of the Green's function:
\begin{equation}
    \mathscr{G} =
    \mqty(
    \mathscr{G}_{00} & \mathscr{G}_{01} & \cdots & \mathscr{G}_{0,N+1}
    \\
    \mathscr{G}_{10} & \mathscr{G}_{11} & \cdots & \mathscr{G}_{1,N+1}
    \\
    \vdots & \vdots & \ddots & \vdots
    \\
    \mathscr{G}_{N+1,0} & \mathscr{G}_{N+1,1} & \cdots & \mathscr{G}_{N+1,N+1}
    ).
\end{equation}
From the Landauer formula~\cite{Landauer1957}, the electrical conductance $G$ in the out-of-plane direction is obtained by
\begin{equation}
    G(E,\vb*{k}_{\para}) = \frac{2e^2}{h} \sum_{\mu\nu} |t_{\mu\nu}|^2,
     \label{eq:G_of_k}
\end{equation}
where the factor $2$ is for the spin degree of freedom.
Finally, the total conductivity across the interface per unit area is obtained by
\begin{align}
        g(E)
        &= 
        \frac{1}{S_{\mathrm{tot}}}
        \sum_{\vb*{k}_{\para} \in \mathrm{MBZ}} G(E,\vb*{k}_{\para})
\nonumber\\
&=
\frac{1}{(2\pi)^2}
\int_{\rm MBZ} d^2 k_\para \,
    G(E,\vb*{k}_{\para}).
\label{eq:conductivity}
\end{align}
where $S_{\mathrm{tot}}$ is the total cross section of the system along the $x$-$y$ plane.
The contact resistivity of the twisted interface is given by $1/g(E)$.
}

{
We can also calculate LDOS in the middle part from the Green's function.
The LDOS of the $n$-th slice is obtained as
\begin{equation}
    \rho(E,\vb*{k}_{\para},n) = -\frac{1}{\pi}
    \mathrm{Im} \, \mathrm{tr} \,
    \mathscr{G}_{nn},
\label{eq:ldos}
\end{equation}
where the trace sums up sublattice or orbital degrees freedom and also the in-plane wavenumbers to span the Hamiltonian matrix.
In twisted graphite, a slice is composed of two graphene monolayers. 
To calculate the LDOS of each layer, we can restrict the summation over  sublattices $(A,B,A',B')$ in the trace in Eq.\eqref{eq:ldos} to just $(A,B)$ or $(A',B')$.
}


\section{Electrical conductivity in twisted graphite interface}
\label{sec:results}


\begin{figure}
\begin{center}
   \includegraphics [width=0.9\linewidth]{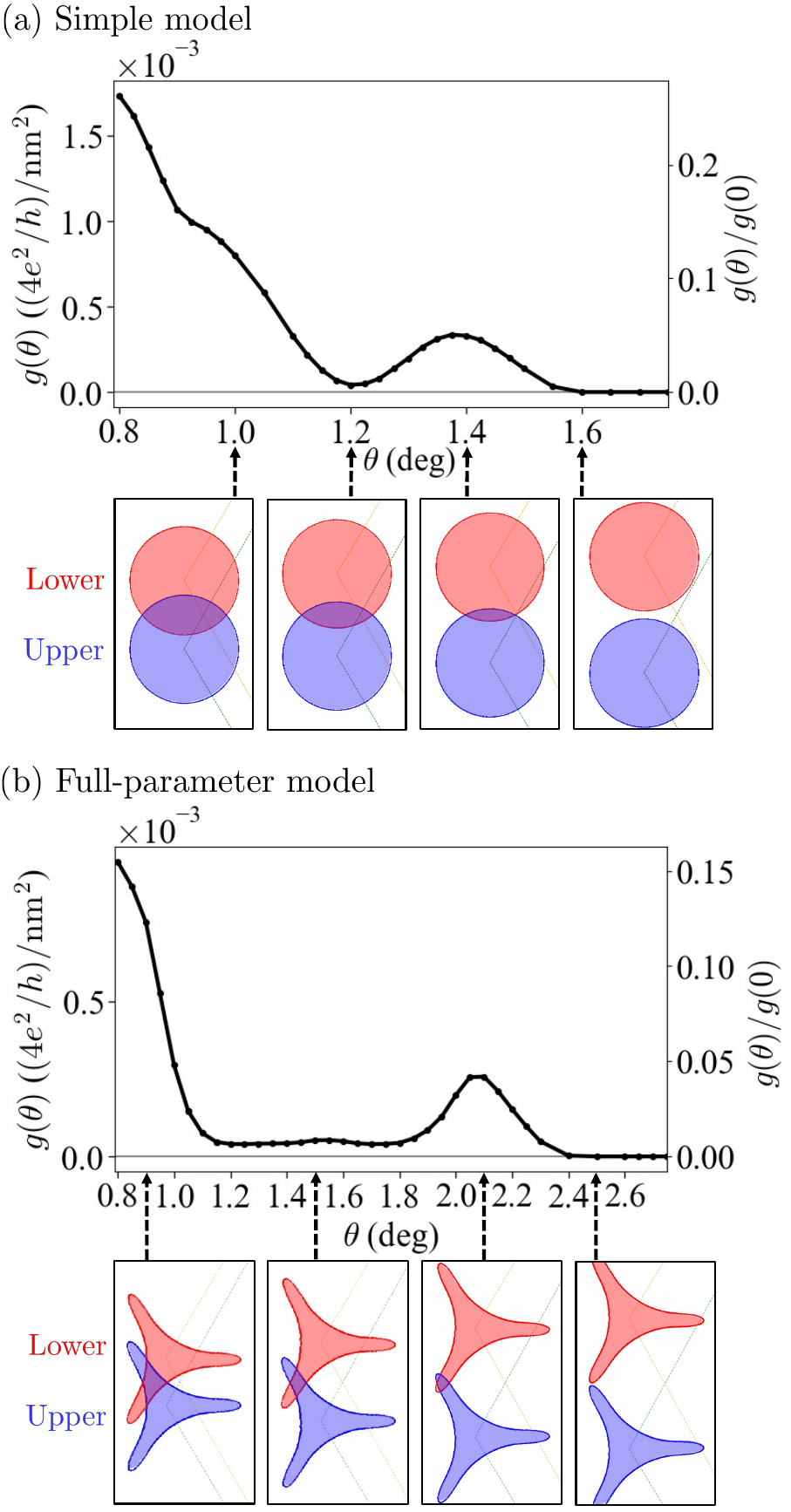}
   \caption{
 (a) Perpendicular electrical conductivity as a function of the twist angle $\theta$, in the simple model of twisted graphite. The lower panels show the Fermi surfaces projected onto $k_xk_y$ plane, for the lower (red) and upper (blue) bulk graphites with $\theta=1.0^{\circ}$, $1.2^{\circ}$, $1.4^{\circ}$, $1.6^{\circ}$.
   (b) Similar plots for the full-parameter model. The Fermi surfaces are shown for $\theta=0.9^{\circ}$, $1.5^{\circ}$, $2.1^{\circ}$, $2.5^{\circ}$.
            }\label{fig:result}
 \end{center}
 \end{figure}

We calculate the perpendicular electrical conductivity 
of twisted graphite by applying 
Eq.~\eqref{eq:conductivity}
to the Hamiltonian Eq.~\eqref{eq:overall-Hamil3}, with no middle part.
Figure~\ref{fig:result} shows the conductivity
as a function of the twist angle $\theta$,
calculated for (a) the simple graphite model and (b) the full parameter model (see Sec.~\ref{subsec:bulk-graphite}).
Here we take the Fermi energy at the charge neutral point, $E=0$. 
 The vertical axis is scaled by $4e^2/h/{\rm nm}^2$ on the left, and also by $g(\theta=0)$ on the right.
{ The $g(\theta=0)$ represents the ballistic conductance of bulk 3D graphite,
 and $g(\theta)/g(0)$ can be regarded as an effective transmission coefficient.
 }

{
In the simple model [Fig.~\ref{fig:result}(a)], the conductivity decreases when the twist angle is increased, and it completely vanishes in $\theta > 1.6^\circ$.
This tendency can be understood in terms of the overlap of the Fermi surfaces.
By a twist, the Brillouin zones of the upper and lower graphite are rotated by $\pm\theta/2$ [see Fig.~\ref{fig:MBZ}(b)],
and then the in-plane projections of respective Fermi surfaces are separated as shown by blue and red filled circles
in Fig.~\ref{fig:result}(a).
Obviously, the overlapped region of the Fermi surface projections is diminished with increase of the angle, and the conductivity drops accordingly.
For $\theta > 1.6^\circ$, the two Fermi surfaces are completely separated and the conductivity goes to zero.
Notably, we observe a non-monotonic behavior in the range of $1.2^\circ < \theta < 1.4^\circ$,
where the conductivity reaches nearly zero
at $\theta\approx1.2^\circ$, and it takes a peak at $\theta\approx1.4^\circ$.
This cannot be simply explained by the Fermi surface overlap, which just monotonically decreases in increasing the twist angle.
}

{
We also see a similar behavior
in the full-parameter model as well, as shown in Fig.~\ref{fig:result}(b).
With the increase of the twist angle, the conductance drops down to $\theta \approx 1^\circ$,
while it recovers and peaks at $\theta \approx 2.1^\circ$.
It finally vanishes in $\theta>2.4^\circ$ where the Fermi surface overlap disappears.
}
%


\begin{figure}
\begin{center}
   \includegraphics [width=1.\linewidth]{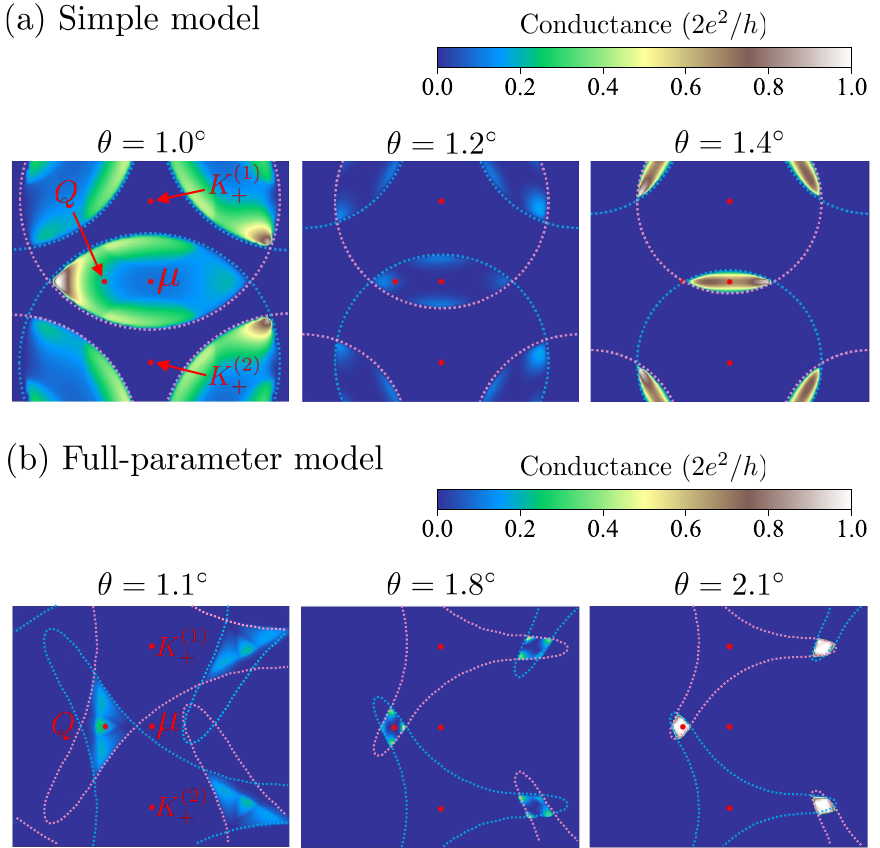}
   \caption{
   Density plot of the momentum-resolved conductance $G(E=0,\vb*{k}_{\para},\theta)$ on $\bm{k}_\para$-space, calculated for (a)~the simple model and (b)~the full-parameter model.
   The magenta (cyan) curve is the outline of the projected Fermi surface of the lower (upper) graphite.
            }\label{fig:conductance}
 \end{center}
 \end{figure}

{
To consider the origin of the dip-and-peak structure, we examine the $k$-resolved conductance $G(E=0,\vb*{k}_{\para})$ defined by Eq.~\eqref{eq:G_of_k}.
Figure \ref{fig:conductance}(a)
shows the density plot of $G(E=0,\vb*{k}_{\para})$ 
on $\vb*{k}_{\para}$-space, calculated for the simple model.
The red (blue) circle represents the outline of the upper (lower) projected Fermi surface.
We observe that the finite amplitude is indeed present only in the overlapping region.
At $\theta = 1.2^\circ$, however, the $k$-resolved conductance is strongly suppressed around the overlap center at the $\mu$ point, and actually
this vanishing amplitude is responsible for the dip of total conductivity $g(\theta)$ at $\theta = 1.2^\circ$ [Fig.~\ref{fig:result}(a)].
In the full-parameter model, a similar decrease of the conductance $G(E=0,\vb*{k}_{\para})$ is found around the overlap center, in the range of $1.2^\circ<\theta<1.9^\circ$ [Fig.~\ref{fig:conductance}(b) for $\theta=1.8^\circ$].
Due to the trigonal warping effect on the Fermi surface, the overlap region is located around the $Q$ point between $\gamma$ and $\mu$ 
[see Fig.~\ref{fig:MBZ}(b)].
}



\begin{figure}
\begin{center}
   \includegraphics[width=0.8\linewidth]{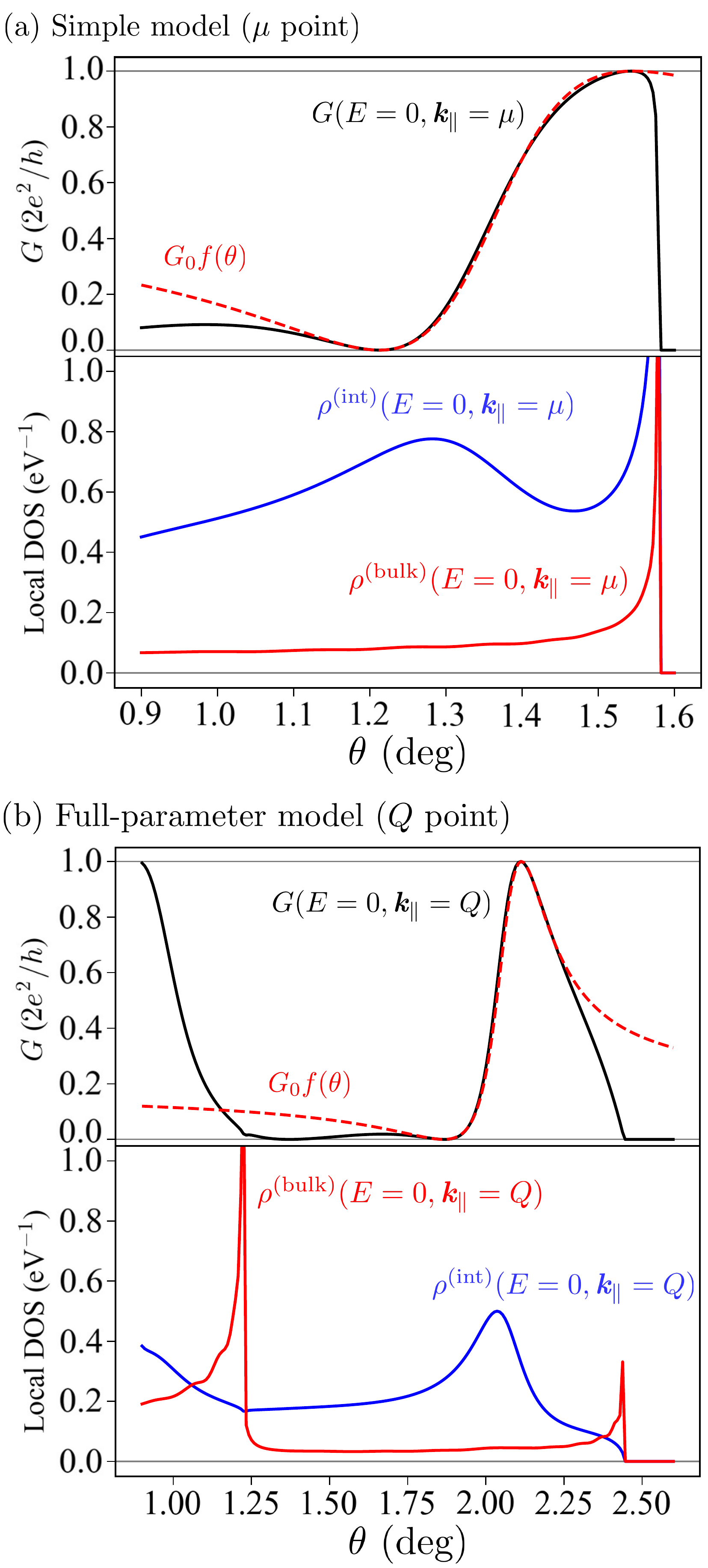}
   \caption{
    (a) Twist-angle dependences of momentum-resolved conductance $G(E,\bm{k}_\para)$
    (upper panel)
    and interface/bulk LDOS 
    $\rho^{(\mathrm{int})/(\mathrm{bulk})}(E,\vb*{k}_{\para})$ (lower panel) in the simple model
    with $E=0$ and $\vb*{k}_{\para} =\mu$.
    The red dashed line in the upper panel represents the fitted Fano funciton
    $G_0 f(\theta)$ (see the text).
    (b) Similar plots for the full-parameter model, where $\vb*{k}_{\para} $ is taken at $Q$.
    }\label{fig:fano}
\end{center}
\end{figure}


\section{Fano resonance by interface-localized state}
\label{sec:discussion}

{
In the following, we demonstrate that the vanishing $k$-resolved conductance $G(E=0,\vb*{k}_{\para})$ argued in the previous section is attributed to the Fano resonance by the interface-localized level.
 We focus on the Fermi surface overlap center, i.e., $\vb*{k}_\para =\mu$ for the simple model and $\vb*{k}_\para = Q$ for the full-parameter model,
and plot the conductance $G(E=0,\vb*{k}_\para)$ against the twist angle $\theta$. The results are shown in the upper panels of Figs.~\ref{fig:fano}(a) and \ref{fig:fano}(b)
for simple and full parameter models, respectively.
In both cases, the conductance exhibits a sort of resonant behavior, where $G/(2e^2/h)=1$ and 0 in Fig.~\ref{fig:fano} can be viewed as
resonant and anti-resonant points, respectively.
Here the maximum of the conductance is $G=2e^2/h$,
since the number of conducting channels per spin 
of the non-twisted regions is 1 in the angle range of the figure. 
}

{
The result implies that a resonance occurs between the bulk state of graphite and a interface-localized state.
To identify associated interface states, we calculate the LDOS at the interface and also in the bulk region
by using Eq.~(\ref{eq:ldos}).
In the numerical calculation, we introduce a finite middle region (Fig.~\ref{fig:3d_system_structure}) containing $N=50$ unit cells (100 graphene layers).
We define the interface/bulk LDOS by
\begin{align}
& \rho^{(\mathrm{int})}(E,\vb*{k}_{\para})
    =
    -\frac{1}{\pi}
    \mathrm{Im} \, \mathrm{tr}^{(\mathrm{int})} \,
    \mathscr{G},
\\
&   \rho^{(\mathrm{bulk})}(E,\vb*{k}_{\para})
    =
    -\frac{1}{\pi}
    \mathrm{Im} \, \mathrm{tr}^{(\mathrm{bulk})} \,
    \mathscr{G}.
\end{align}
Here $\mathrm{tr}^{(\mathrm{int})}$ stands for trace over the wave bases belonging to the top graphene layer of the lower graphite, and the bottom layer of the upper graphite.
The $\mathrm{tr}^{(\mathrm{bulk})}$ runs over the complementary bases 
in the middle part, which are not included in $\mathrm{tr}^{(\mathrm{int})}$.
}

{
The lower panel of Fig.~\ref{fig:fano}(a) shows the angle dependence of  $\rho^{\mathrm{(int)}}$ and  $\rho^{\mathrm{(bulk)}}$  with $E=0$ and $\vb*{k}_\parallel = \mu$ in the simple model.
We normalized $\rho^{\mathrm{(bulk)}}$ to the value per two graphene layers,
to be directly compared with $\rho^{\mathrm{(int)}}$.
The $\rho^{\mathrm{(int)}}$ exhibits a broad peak centered at $\theta \approx 1.3^\circ$, which indicates the emergence of an interface-localized state.
For the full-parameter model, similarly, a peak of $\rho^{\mathrm{(int)}}$  appears at $\theta \approx 2.0^\circ$ as seen in Fig.~\ref{fig:fano}(b).
In both models, the bulk LDOS $\rho^{\mathrm{(bulk)}}$  does not show a peak at the corresponding positions. 
}

{
Generally, a system with weakly-coupled continuum and discrete spectra shows the Fano resonance~\cite{Fano1961},
which is characterized by an asymmetric line shape in the system's response.
In twisted graphite, the emergent interface-localized state is considered to be a discrete state, while the bulk state contributes to a continuum spectrum.
The transmission probability should then be given by the Fano function as a function of the twist angle,
\begin{equation}
    f(\theta) = \frac{(x+q)^{2}}{1+x^2},
    \quad 
    x=\frac{2(\theta-\theta_{0})}{\Delta\theta},
\end{equation}
where 
$\Delta\theta$ represents
the broadening of the discrete state,
$\theta_0$ is approximately equal to the position of the discrete state,
and $q$ determines the asymmetry of the peak.
}


{
Here we fit the conductance curve $G(\theta)$ by the Fano function $G_0 f(\theta)$, and plot the obtained curves with red dashed lines in Figs.~\ref{fig:fano}(a) and \ref{fig:fano}(b).
We employ the parameters 
$\theta_0 = 1.357$, $\Delta\theta = 0.330$, $q = 0.877$ for the simple model, and $\theta_0 = 2.070$, $\Delta\theta = 0.190$, $q = 2.156$
for the full-parameter model.
The amplitude parameter is taken as $G_0=(2e^2/h)/(1+q^2)$ for the correct peak height.
In both cases, we see a fairly nice fitting with $G(\theta)$ in a wide range of the twist angle. Also the obtained $\theta_0$ nearly coincides with the peak position of the interface LDOS-curve, which corresponds to the position of the discrete level.
Regarding this good agreement, we conclude that the dip-and-peak structure of the perpendicular conductivity is attributed to the Fano resonance.
}


\begin{figure*}
\begin{center}
   \includegraphics [width=0.8\linewidth]{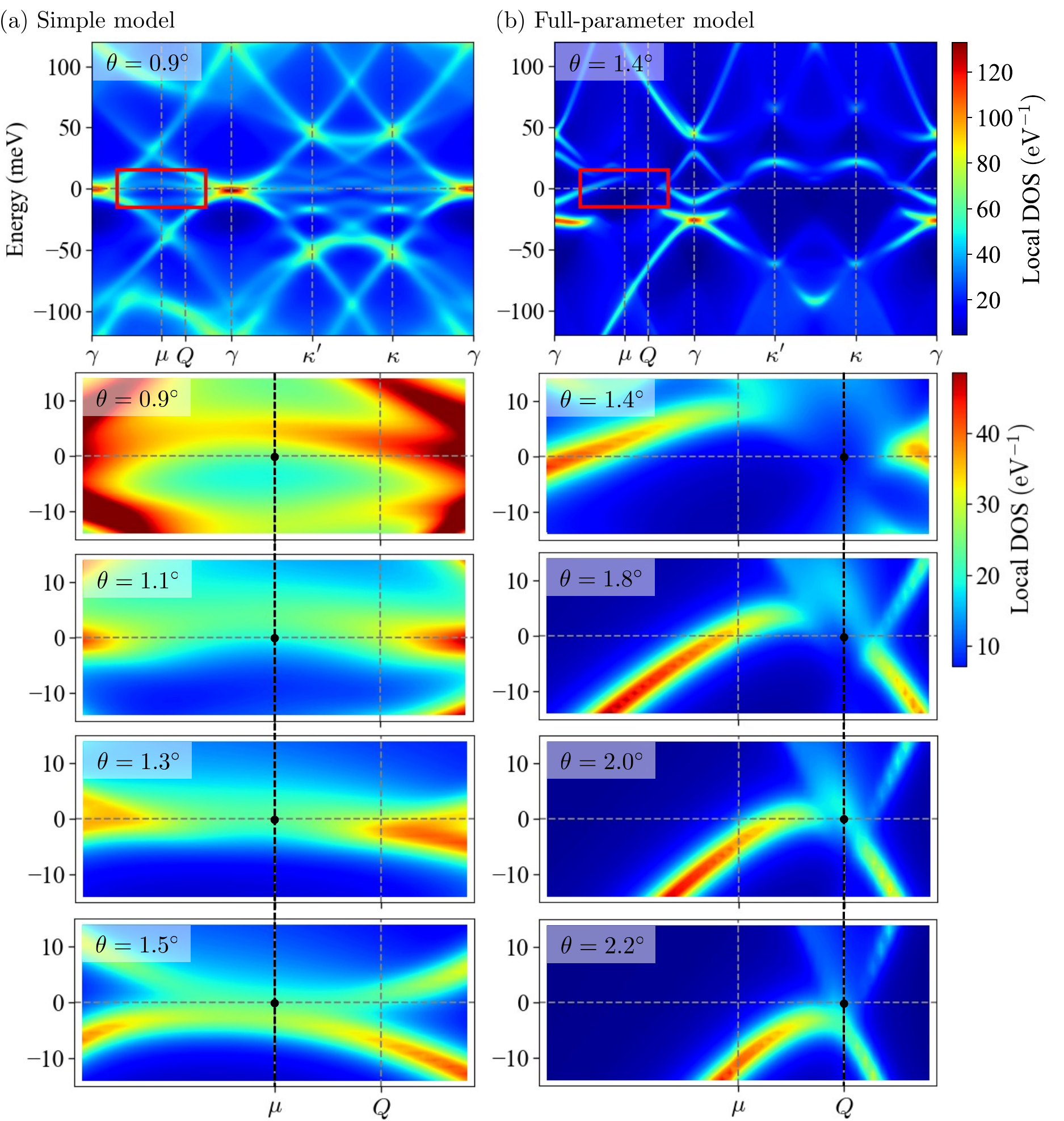}
   \caption{
   Density plots of the interface LDOS $\rho^{(\mathrm{int})}(E,\vb*{k}_{\para})$ in (a) the simple and (b) the full-parameter model.
   Bright traces indicate the interface-localized bands.
 In each figure, the top panel shows 
   a wide-range plot for a typical twist angle $\theta =0.9^\circ$, $1.4^\circ$ for (a) and (b), respectively.
   The lower panels are magnified plots in the low-energy region indicated by a red rectangle in the top panels, calculated for several $\theta$'s.
            }\label{fig:ldos}
 \end{center}
 \end{figure*}
 
{
Finally, we demonstrate that the interface-localized state, which induces the Fano resonance, is a manifestation of a flat-band-like structure inherited from TBG.
Here we compute the interface LDOS $\rho^{(\mathrm{int})}(E,\vb*{k}_{\para})$ over a wide range of energy and momentum to identify the energy band associated with the interface-localized state.
In the top panel of Fig.~\ref{fig:ldos}(a), we show the density plot of $\rho^{(\mathrm{int})}(E,\vb*{k}_{\para})$ in the simple model with $\theta=0.9^\circ$.
A spectral broadening $\delta = 3 \, \mathrm{meV}$ is introduced for the sake of visibility.
Most of bright lines observed in the figure 
do not appear in the bulk LDOS (not shown), 
indicating that these lines correspond to the interface bands.
The nearly-horizontal branch located at the energy of $E=0$ is the remnant of the flat band. Indeed, if we ignore the hopping parameters $\gamma_2$ and $\gamma_5$
retaining only $\gamma_0$ and $\gamma_1$,
the band becomes perfectly flat at $E=0$~\cite{Cea2019}.
}

{
The second top panel in Fig.~\ref{fig:ldos}(a) shows a magnified plot in the low-energy region, indicated by the red rectangle in the top panel. The third and lower panels are corresponding plots for different twist angles. 
When we increase the twist angle $\theta$, the nearly flat band is lowered and traverses the point 
$(\vb*{k}_{\para},E) = (\mu, 0)$ (indicated by a black dot)  around $\theta=1.3^\circ$. This corresponds to the interface LDOS peak in Fig.~\ref{fig:fano}(a).
Figure \ref{fig:ldos}(b) presents similar plots for the full-parameter model.
We observe that the flat band obtains some dispersion, yet it still persists in the low-energy region indicated by the red rectangle. 
In increasing the twist angle, 
the high-amplitude part crosses the
$(\vb*{k}_{\para},E) = (Q, 0)$ point,
and this causes the interface LDOS peak in Fig.~\ref{fig:fano}(b).
}

\section{Conductivity peak near 21.8$^\circ$}
\label{sec:commensurate}

\begin{figure}
\begin{center}
   \includegraphics [width=0.7\linewidth]{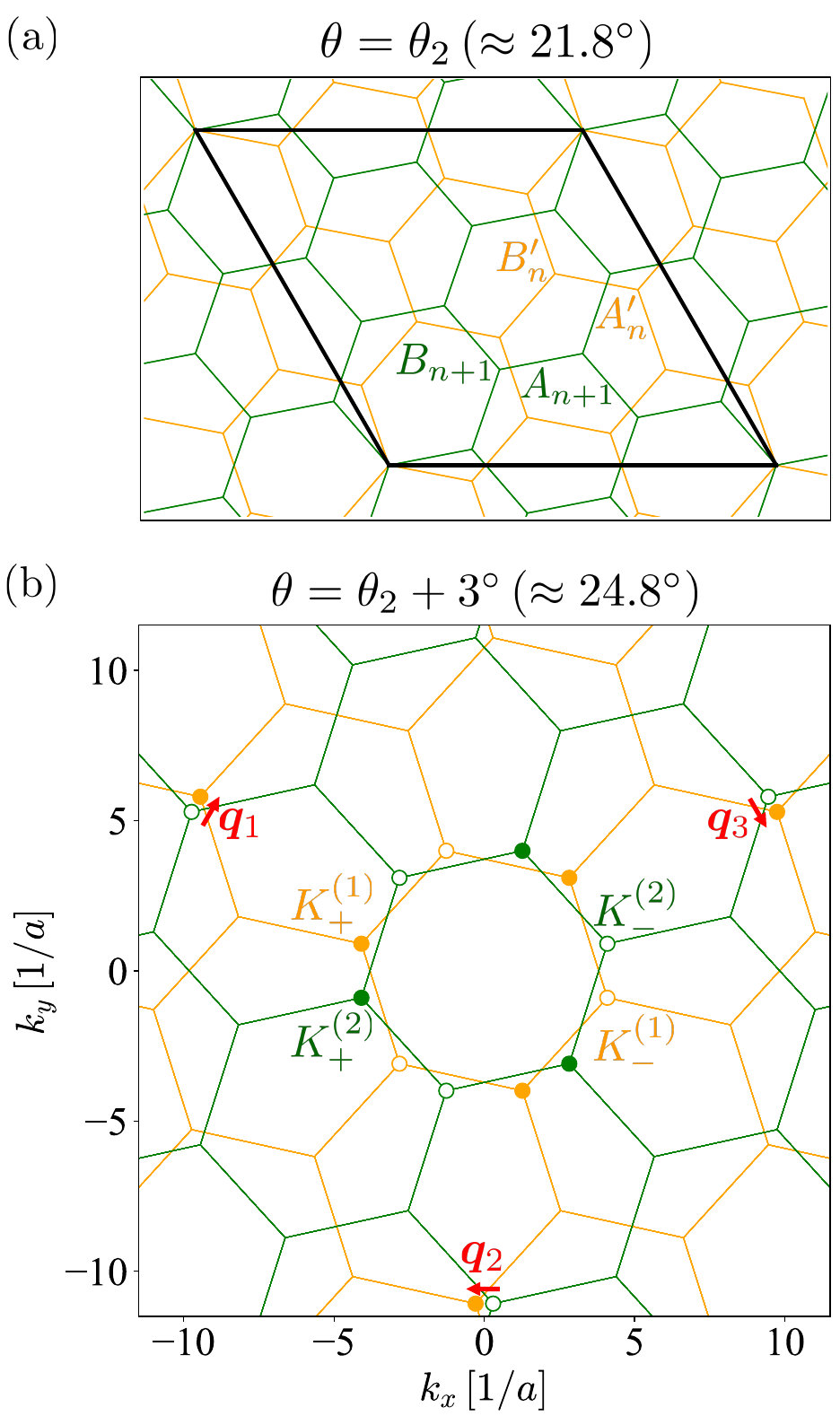}
   \caption{
   (a) Interface atomic structure in a twisted graphite at the commensurate angle $\theta=\theta_2 \approx 21.8^\circ$.
   The yellow (green) hexagons are the topmost (bottommost) layer of the lower (upper) graphite. 
   The $\sqrt{7}\times\sqrt{7}$ unit cell is represented by the black diamond.
   (b) Brillouin zone of the lower (yellow) and upper (green) honeycomb lattice in a slightly misaligned angle $\theta= \theta_2 + 3^\circ \, (\approx 24.8^\circ)$. 
   The coupling wavenumbers $\vb*{q}_1,\vb*{q}_2,\vb*{q}_3$ are shown in the same manner as the low-angle case in Fig.~\ref{fig:MBZ}(b).
   }\label{fig:lattice_21.8}
 \end{center}
 \end{figure}

While we considered the small twist angle regime $\theta \approx 0^\circ$ in the preceding sections,
the coherent interlayer transport occurs also at other commensurate angles.
In this section, we examine the perpendicular conductivity near the second commensurate angle $\theta= \theta_2 \approx 21.8^\circ$, where the honeycomb lattices become exactly periodic with the $\sqrt{7}a\times\sqrt{7}a$ unit cell as shown in Fig.~\ref{fig:lattice_21.8}(a)~\cite{Shallcross2008,Mele2010,Shallcross2010}.
The interlayer transport in twisted graphene layers
was theoretically investigated in the incoherent transport regime, and the conductivity enhancement at the commensurate angles was predicted \cite{Bistritzer2010}.
Recent experiments reported a sharp conductance peak at $\theta= \theta_2$ in variable-angle graphite devices \cite{Koren2016, Li2018, Inbar2023}.
In the following, we describe the qualitative angle-dependent behavior near $\theta_2$, by employing the same theoretical approach adopted in the previous section.


The interlayer coupling across the twisted interface near $\theta_2$ can be captured by a similar model to Eq.~\eqref{eq:overall-Hamil3}  for $\theta\approx 0^\circ$.
Figure \ref{fig:lattice_21.8}(b)
shows the $k$-space diagram for
$\theta = 24.8^\circ$, which is near $\theta_2$,
where yellow and green honeycomb lattices represent the extended Brillouin zones for lower and upper graphite.
As in Fig.~\ref{fig:MBZ}(b),
we define $\vb*{q}_1, \vb*{q}_2, \vb*{q}_3$  as the smallest separations between the corner points of upper and lower layers. Note that the $\bm{q}_i$'s vanish at $\theta =\theta_2$.
These three vectors determine the moir\'e superperiod when $\theta$ is slightly away from $\theta_2$,
giving the smallest Fourier components in the interlayer Hamiltonian  \cite{Fujimoto2022,Koshino2015}.

When the lattice relaxation is neglected,
the interlayer coupling magnitude associated with the $\bm{q}_i$'s is given by $t(\bm{Q})$, where $\bm{Q}$ is the $k$-space position of the corresponding corner point, and the function $t(\bm{k})$ is the Fourier transform of the interlayer hopping amplitude \cite{Bistritzer2010, Fujimoto2022,Koshino2015}.
In the case of $\theta \approx 0^\circ$ [Fig.~\ref{fig:MBZ}(b)], $\bm{Q}$ is the $\bm{K}_{\xi}$ point,
giving the coupling magnitude of $t(K)$
with $K =|\bm{K}_\xi|= 4\pi/(3a)$.
This leads to the simplest interlayer coupling Hamiltonian
for TBG, which is Eq.~\eqref{eq:interlayer} with
$u$ and $u'$ replaced by $t(K)$ \cite{Bistritzer2011}.
Note that the difference between $u$ and $u'$ in  Eq.~\eqref{eq:interlayer} is introduced to effectively describe the lattice relaxation, which is only effective in the small $\theta$ regime.

In $\theta \approx \theta _2$, the corner points $\bm{Q}$ are located at distance $\sqrt{7}K$
from the origin, 
resulting in an interface coupling matrix of  \cite{Koshino2015},
\begin{equation}
\begin{split}
    &T_{\mathrm{int}}^{2\times2}(\vb*{r}) =
    \\
    &t(\sqrt{7}K) \left[
    \mqty(1 && \omega \\ \omega^{*} && 1)
    e^{i\vb*{q}_{1} \cdot \vb*{r}} +
    \mqty(1 && 1 \\ \omega && \omega)
    e^{i\vb*{q}_{2} \cdot \vb*{r}} +
    \mqty(1 && \omega^{*} \\ 1 && \omega^{*})
    e^{i\vb*{q}_{3} \cdot \vb*{r}}
    \right].
\label{eq:commensurate_coupling}
\end{split}
\end{equation}
The coupling amplitude $t(\sqrt{7}K)$ is much smaller than $t(K)$ for $\theta\approx 0^\circ$,
as the Fourier transform $t(k)$ is a decaying function.
While the value of this factor strongly depends on the details of the model \cite{Mele2010,Bistritzer2011,Koshino2015},
here we employ $t(\sqrt{7}K) = 1.3 \, \mathrm{meV}$,
which is extracted from the tight-binding hopping model fitted to the LDA calculation \cite{Perebeinos2012,Habib2013,Koren2016}.
In Appendix~\ref{app:21.8}, we evaluate $t(\sqrt{7}K)$ by an alternative approach using the first-principles band calculation for the commensurate TBG of $\theta =\theta_2$,
to obtain the parameter of the same order.


\begin{figure}
\begin{center}
   \includegraphics [width=0.9\linewidth]{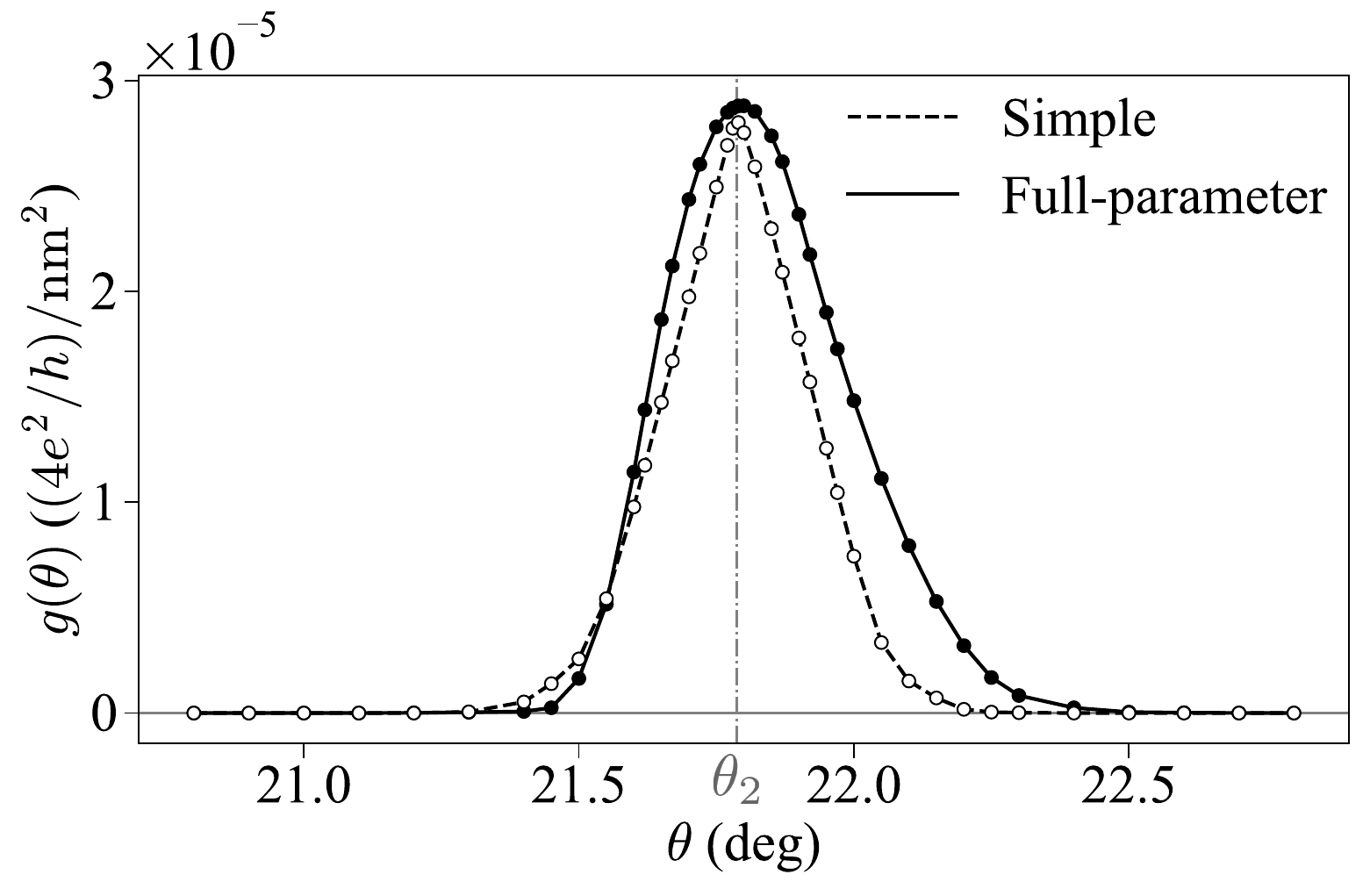}
   \caption{
   Perpendicular conductivity around $\theta=\theta_2 \approx 21.8^\circ$ in the simple model (dashed curve) and full-parameter model (solid curve).
   The interlayer coupling is taken as $t(\sqrt{7}K) = 1.3~\mathrm{meV}$.
            }\label{fig:comm_cond}
 \end{center}
 \end{figure}

Another important difference from $\theta \approx 0^\circ$ is that the twisted interface of 
$\theta \approx \theta_2$ hybridizes electronic states at the opposite valleys, 
$K_\xi^{(1)}$ and $K_{-\xi}^{(2)}$,
as seen in Fig.~\ref{fig:lattice_21.8}(b).
As a result, pseudo-spin chirality of the Bloch electron is inverted between the two layers,
and $H^{(l)}$ and $T^{(l)}$ in Eq.~\eqref{eq:overall-Hamil3}
are replaced by
\begin{align}
& H^{(1)} = H_\xi[(R^{(1)})^{-1}\bm{k}_{\para}],
\quad
 T^{(1)} = T_\xi[(R^{(1)})^{-1}\bm{k}_{\para}],
 \nonumber\\
& H^{(2)} = H_{-\xi}[(R^{(2)})^{-1}\bm{k}_{\para}],
\quad
 T^{(2)} = T_{-\xi}[(R^{(2)})^{-1}\bm{k}_{\para}].
\label{eq:twistedgr-Tmatrix_21.8}
\end{align}


By using the Hamiltonian Eq.~\eqref{eq:overall-Hamil3} with Eqs.~\eqref{eq:commensurate_coupling} and
\eqref{eq:twistedgr-Tmatrix_21.8},
we calculate the electrical conductivity in the same manner as in the small-angle cases.
The resulting contact conductivity $g(\theta)$ near $\theta=\theta_2\approx 21.8^\circ$ is shown in Fig.~\ref{fig:comm_cond}.
For both simple and full-parameter models, we see that $g(\theta)$ peaks around the commensurate angle $\theta_2$.
Since the interlayer coupling is perturbative, the multiple interlayer-scattering processes in the Green's function are negligible, so that the transmission is dominated by the first-order hopping process.
Therefore, $t_{\mu\nu}$ is approximately proportional to $t(\sqrt{7}K)$, leading to the relationship $g \propto t(\sqrt{7}K)^2$.
{It should be noted that,  in this high twist angle regime, no interface-localized states appear near $E=0$, and hence the resonant behavior does not occur in the conductivity unlike in the low-angle regime. The sharp conductance peak at $\theta = 21.8^\circ$ is qualitatively explained by the Fermi surface overlap at the remote $k$ point in Fig.~\ref{fig:lattice_21.8}(b).}

The commensurate conductance peak at $\theta \approx 21.8^\circ$ was experimentally observed in twisted interfaces between graphite and graphite 
\cite{Koren2016}, graphite and graphene \cite{Chari2016}, and graphene and graphene \cite{Inbar2023}.
Our calculation in Fig.~\ref{fig:comm_cond} roughly reproduces the order of magnitude of 
the interface conductivity in the graphite-graphite device \cite{Koren2016}, which is close to our situation.
It should be noted that the interface transport in real devices  has also a considerable contribution from the phonon-mediated hopping,
which gives a smooth background mildly depending on the angle \cite{Perebeinos2012, Koren2016}.

\section{Conclusion}
\label{sec:conclusion}

We have developed a theoretical method to describe the transport in twisted 3D systems by using the recursive Green's function approach. By using the formulation, we calculated the perpendicular conductivity in the twisted graphite.
The calculated conductivity exhibits a nonmonotonic dip-and-peak structure against the twist angle, due to vanishing transmission at the overlap center of the Fermi surfaces.
By examining the LDOS spectrum, we revealed that
the drop of the conductance is caused by the Fano resonance between the bulk state and the interface-localized state, which is the remainder of a flat band of TBG.
We also calculated the perpendicular conductivity at the twist angles around the second commensurate angle $\theta \approx 21.8^\circ$, 
and find a sharp peak around $21.8^\circ$, which is consistent with experimental observation~\cite{Koren2016,Inbar2023}.

{Although we limited our argument to the twisted junction of two graphite pieces in this paper, the proposed formulation is applicable to diverse twisted systems.
For instance, we can consider a system composed of two twisted interfaces with an $N$-layer graphite in the middle section.
There we anticipate the emergence of localized states in the middle section
depending on its thickness $N$, leading to complex resonances in the out-of-plane conductance through a similar mechanism.
We can also extend our analysis to systems incorporating numerous twisted interfaces, including 3D graphite spirals \cite{Cea2019,Wang2023} and alternating twisted multilayer graphenes \cite{Burg2022,Nguyen2022}. 
This applies not only to graphitic materials; the formulation can be effectively extended to explore twisted interfaces of diverse metallic and superconducting materials. 
Applying the present method to study the perpendicular electronic transport in these systems would be intriguing future research.
}


\appendix
\section{Recursive Green's function method}
\label{app:recursive}
In the present appendix, we explain the recursive Green's function method~\cite{Ando1991,Lewenkopf2013} to calculate the Green's function of Eq.~(\ref{eq:reduced_hamil}), $\mathscr{G} = (E-\mathcal{H}_{\mathrm{reduced}})^{-1}$.
Each block of the Green's function $\mathscr{G}_{nm}$ can be calculated from lower (upper) Green's functions $\mathscr{G}^{(1)}_{nm}$ ($\mathscr{G}^{(2)}_{nm}$), as shown below.
Starting from the lower bulk Green's function $g^{(1)} = (E - H^{(1)} - \Sigma^{(1)})^{-1}$, the lower Green's functions are computed by recursive relations
\begin{equation}
    \begin{split}
        \mathscr{G}^{(1)}_{11} &= (E-h_{1}-T_{10}g^{(1)}T_{01})^{-1},
        \\
        \mathscr{G}^{(1)}_{01} &= g^{(1)} T_{01} \mathscr{G}^{(1)}_{11},
        \\
        \mathscr{G}^{(1)}_{nn} &= (E-h_{n}-T_{n,n-1}\mathscr{G}^{(1)}_{n-1,n-1}T_{n-1,n})^{-1},
        \\
        \mathscr{G}^{(1)}_{0n} &= \mathscr{G}^{(1)}_{0,n-1}T_{n-1,n}\mathscr{G}^{(1)}_{nn},
    \end{split}
\label{eq:GL}
\end{equation}
where $2 \leq n \leq N$.
Similarly, from the upper bulk Green's function $g^{(2)} = (E - H^{(2)} - \Sigma^{(2)})^{-1}$, the upper Green's functions can be gained by equations
\begin{equation}
    \begin{split}
        \mathscr{G}^{(2)}_{NN} &= (E-h_{N}-T_{N,N+1}g^{(2)}T_{N+1,N})^{-1},
        \\
        \mathscr{G}^{(2)}_{N+1,N} &= g^{(2)}T_{N+1,N}\mathscr{G}^{(2)}_{NN},
        \\
        \mathscr{G}^{(2)}_{nn} &= (E-h_{n}-T_{n,n+1}\mathscr{G}^{(2)}_{n+1,n+1}T_{n+1,n})^{-1},
        \\
        \mathscr{G}^{(2)}_{N+1,n} &= \mathscr{G}^{(2)}_{N+1,n+1}T_{n+1,n}\mathscr{G}^{(2)}_{nn},
    \end{split}
\label{eq:GR}
\end{equation}
where $1 \leq n \leq N-1$.
Finally, we can obtain the full Green's function $\mathscr{G}$ by recursions
\begin{equation}
    \begin{split}
        \mathscr{G}_{nn} &= (E - h_{n} - T_{n,n-1}\mathscr{G}^{(1)}_{n-1,n-1}T_{n-1,n} 
        \\ &\hspace{2cm} - T_{n,n+1}\mathscr{G}^{(2)}_{n+1,n+1}T_{n+1,n})^{-1},
        \\
        \mathscr{G}_{n-1,n} &= \mathscr{G}^{(1)}_{n-1,n-1}T_{n-1,n}\mathscr{G}_{nn},
        \\
        \mathscr{G}_{0n} &= \mathscr{G}^{(1)}_{0,n-1}T_{n-1,n}\mathscr{G}_{nn},
        \\
        \mathscr{G}_{N+1,n} &= \mathscr{G}^{(2)}_{N+1,n+1}T_{n+1,n}\mathscr{G}_{nn},
        \\
        \mathscr{G}_{n,n+1} &= \mathscr{G}_{nn}T_{n,n+1}\mathscr{G}^{(2)}_{n+1,n+1},
    \end{split}
\label{eq:Gfull}
\end{equation}
where $2 \leq n \leq N-1$.

\section{Alternative estimation of interlayer coupling $t(\sqrt{7}K)$}
\label{app:21.8}
The interlayer coupling parameter $t(\sqrt{7}K)$, which determines the magnitude of the conductivity at the commensurate angle $\theta =\theta_2 (\approx 21.8^\circ)$,
is highly dependent on the detail of a model under consideration.
In Sec.~\ref{sec:commensurate}, we adopted the value of $t(\sqrt{7}K) = 1.3 \, \mathrm{meV}$, which is extracted from the tight-binding hopping model fitted to the LDA calculation \cite{Perebeinos2012,Habib2013,Koren2016}.
Here, we give an alternative evaluation of the parameter based on
the band calculation of TBG at $\theta=\theta_2$ in two different methods,
the effective continuum model and the density functional theory (DFT).

In the continuum model,
we can calculate the energy bands of TBG with $\theta\approx\theta_2$  by the following $4\times4$ Hamiltonian
\begin{equation}
    H^{\mathrm{TBG}} =
    \mqty(
    -\hbar v (k_x,k_y)\cdot\vb*{\sigma} & T_{\mathrm{int}}^{2\times2\dagger}(\vb*{r}) \\
    T_{\mathrm{int}}^{2\times2}(\vb*{r}) & -\hbar v (-k_x,k_y)\cdot\vb*{\sigma}
    ).
\end{equation}
Here the upper and lower diagonal $2\times2$ blocks are the Dirac Hamiltonian of monolayer graphene at $K_+$ (lower layer) and $K_-$ (upper layer), respectively.
Here, note that the interface of $\theta=\theta_2$ hybridizes opposite valleys as explained in Sec.~\ref{sec:commensurate}.
$\vb*{\sigma} = (\sigma_x, \sigma_y)$ is the set of the Pauli matrices.

\begin{figure*}
\begin{center}
   \includegraphics [width=0.6\linewidth]{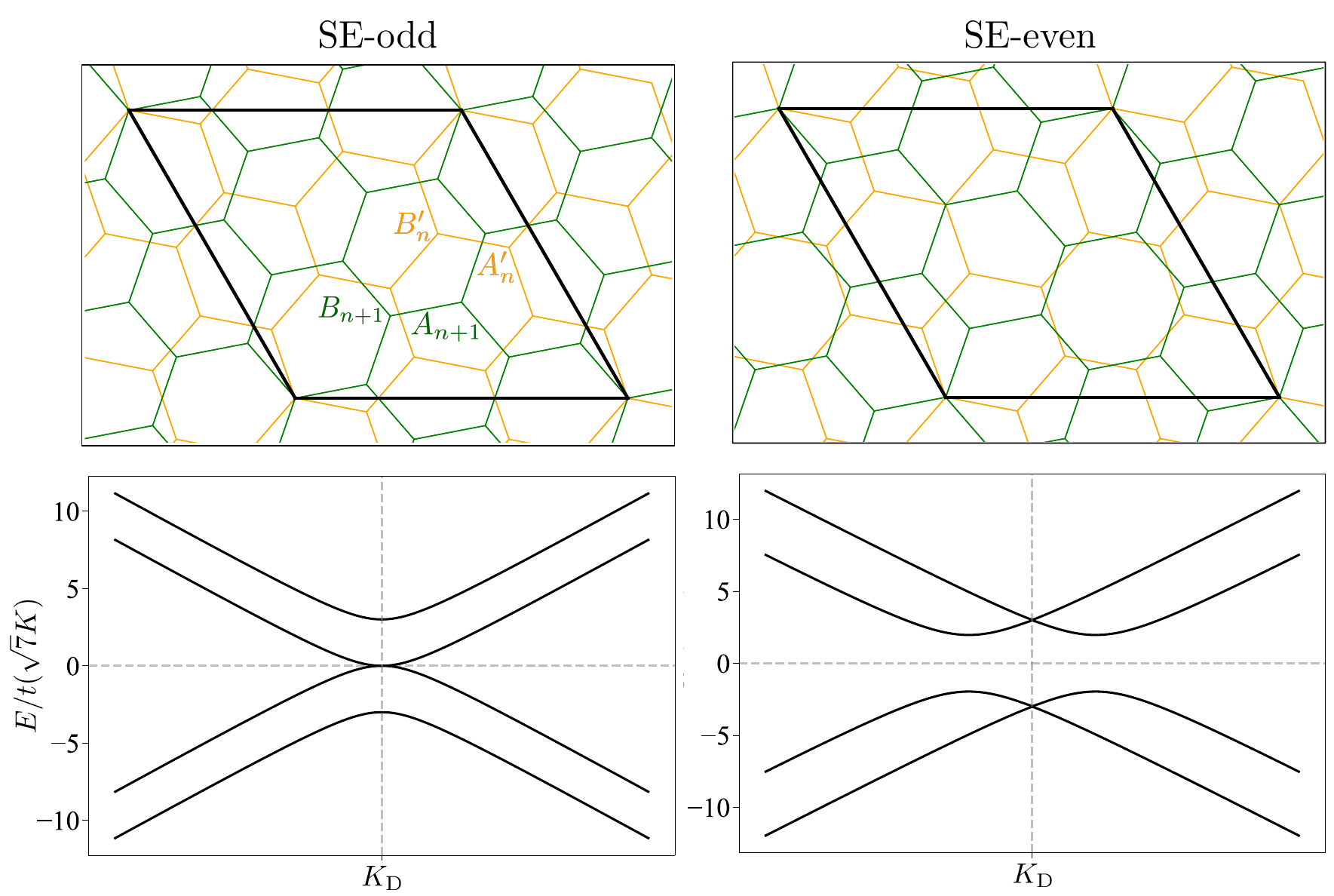}
   \caption{
   Atomic structure and the energy band of the twisted bilayer graphenes of $\theta=\theta_2 \approx 21.8^\circ$ with the SE-odd (left) and SE-even (right) configurations.
   The energy in the band dispersion is scaled by $t(\sqrt{7}K)$.
            }\label{fig:SE_band}
 \end{center}
 \end{figure*}

The off-diagonal block  $T_{\mathrm{int}}^{2\times2}(\vb*{r})$ is the position-dependent interlayer potential given by Eq.~\eqref{eq:commensurate_coupling}.
When the twist angle is slightly shifted from $\theta_2$, the interlayer coupling $T_{\mathrm{int}}^{2\times2}(\vb*{r})$ slowly modulates as a function of position $\vb*{r}$, where the corresponding moir\'e period $\bm{L}_i^{\mathrm{M}} (i=1,2)$ is defined by $\vb*{L}^{\mathrm{M}}_{i} \cdot \vb*{G}^{\mathrm{M}}_{j} = 2\pi\delta_{ij}$
with $\vb*{G}_1^{\mathrm{M}}=\vb*{q}_2-\vb*{q}_1$ and $\vb*{G}_2^{\mathrm{M}}=\vb*{q}_3-\vb*{q}_2$.
Then, the local Hamiltonian with a fixed $\vb*{r}$ corresponds to a commensurate TBG exactly at $\theta = \theta_2$ with a particular interlayer translation.
In Fig.~\ref{fig:SE_band}, we show the calculated energy bands at 
$\vb*{r} = (-\vb*{L}_1^{\mathrm{M}}+2\vb*{L}_2^{\mathrm{M}})/3$ and
$(\vb*{L}_1^{\mathrm{M}}-2\vb*{L}_2^{\mathrm{M}})/3$, 
which correspond to the SE (sublattice exchange)-odd and SE-even structures, respectively~\cite{Mele2010}.
Here the vertical axes is scaled by $t(\sqrt{7}K)$.

We can directly compare the band structure of Fig.~\ref{fig:SE_band} 
with the corresponding DFT band calculation~\cite{Park2019}.
By comparing the band splitting, 
we obtain $t(\sqrt{7}K) \approx 2.25 \, \mathrm{meV}$, which has the same order as $1.3 \, \mathrm{meV}$ adopted in the main text.
In the case of $t(\sqrt{7}K) = 2.25 \, \mathrm{meV}$, the conductivity shown in Fig.~\ref{fig:comm_cond} is enhanced by a factor of three, approximately,
noting that the conductivity is nearly proportional to $t(\sqrt{7}K)^2$.

\begin{acknowledgements}
This work was supported by JSPS KAKENHI Grants No. JP23KJ1497, No. JP20K14415, No. JP20H01840, No. JP20H00127, No. JP21H05236, and No. JP21H05232, and by JST CREST Grant No. JPMJCR20T3, Japan. 
\end{acknowledgements}

\bibliography{reference}
\end{document}